\pdfoutput=1
\documentclass[pra,twocolumn,showpacs,floatfix]{revtex4-1}
\usepackage{times,amsmath,amssymb,amstext,latexsym,float,graphicx,color,ulem}
\usepackage{hyperref}

\hypersetup{colorlinks=true, citecolor=blue, urlcolor=blue, linkcolor=blue}
\begin{document}

\title{Superfluid-droplet crossover in a binary boson mixture on a ring: \\
Exact diagonalization solutions for few-particle systems in one dimension}

\author{L.~Chergui$^1$}
\email{lila.chergui@matfys.lth.se}
\author{J.~Bengtsson$^1$}
\author{J.~Bjerlin$^1$}
\author{P.~St\"urmer$^1$}
\author{G.~M.~Kavoulakis$^{2,3}$}
\author{S.~M.~Reimann$^1$}

\affiliation{$^1$Division of Mathematical Physics and NanoLund, LTH, Lund University, SE-221 00 Lund, Sweden}

\affiliation{$^2$Hellenic Mediterranean University, PO Box 1939, GR-71004, Heraklion, Greece}

\affiliation{$^3$HMU Research Center, Institute of Emerging Technologies, GR-71004, Heraklion, Greece}

\date{\today}

\begin{abstract} {
We investigate the formation of self-bound quantum droplets in a one-dimensional binary mixture of bosonic atoms, 
applying the method of numerical diagonalization of the full Hamiltonian. The excitation spectra and ground-state pair correlations signal the formation of a few-boson droplet when crossing the region of critical inter-species interactions. }The self-binding affects 
the  rotational excitations, displaying a change in the energy dispersion from negative  curvature, associated with superfluidity in the many-body limit, to a nearly parabolic curvature indicative of rigid body rotation.  
We exploit two global symmetries of the system to further analyze the few-body modes in terms of
transition matrix elements and breathing mode dynamics. The exact results are compared to the usual ad-hoc inclusion of 
higher-order contributions in the extended Gross-Pitaevskii equation, showing a remarkable agreement between the few-body regime and the thermodynamic limit in one dimension. 
\end{abstract}

\maketitle
%%%%%%%%%%%%%%%%%%

\section{Introduction}
Beyond mean-field (BMF) effects in ultra-cold atomic gases have long been suggested as the origin of 
self-bound complexes of bosons, fermions or even boson-fermion mixtures~\cite{Bulgac2002,Bedaque2003,Hammer2004}. 
While BMF corrections originating from three-body effects or quantum fluctuations are typically small,
they may play a significant role in weakly interacting binary Bose gases, where different competing mean-field (MF) contributions to the energy may be tuned to almost cancel each other. As a result small BMF corrections, such as the  Lee-Huang-Yang (LHY) quantum fluctuations~\cite{Lee1957}, become consequential to the state of the system and may lead to the formation of a dilute bosonic droplet~\cite{Petrov2015,Petrov2016} due to differences in scaling with density. Albeit initially predicted for binary bosonic gases, the first observations of such quantum fluctuation-stabilized droplets came from experiments on dipolar condensates
~\cite{Kadau2016,Ferrier-Barbut2016,Ferrier-Barbut2016b,Schmitt2016,Chomaz2016}. 
There, the stabilization through quantum fluctuations in a rather similar scenario 
leads to the formation of spatially elongated self-bound droplets and droplet crystals with shapes governed by the dipolar magnetostriction~\cite{Wachtler2016a,Wachtler2016b,Bisset2016,Baillie2017}. The formation of filaments 
in analogy to a Rosensweig transition in a dipolar condensate~\cite{Kadau2016} was also analyzed in Monte Carlo simulations~\cite{Saito2016,Macia2016,Cinti2017}.
Soon after the discovery of dipolar droplets, the original suggestion of self-bound droplets in binary gases~\cite{Petrov2015} was confirmed by two independent observations  in mixtures of potassium atoms in different hyperfine states~\cite{Cabrera2018,Semeghini2018,Cheiney2018}, and followed-up by experimental studies also on hetero-nuclear mixtures ~\cite{Errico2019,ZGuo2021} reporting an increase in droplet lifetimes. 
Although the extended Gross-Pitaevskii (eGP) approach incorporates BMF quantum fluctuation contributions  in a somewhat ad-hoc manner, it describes many of the experimental findings rather well  (see, {\it e.g.},   Refs.~\cite{Wachtler2016a,Wachtler2016b,Bisset2016,Baillie2017,Edler2017,Salasnich2018,Cidrim2018,Roccuzzo2019,Blakie2020,Pal2020,Lee2021} or the reviews~\cite{Bottcher2021,Luo2021}).  In three dimensions, deviations of critical atom numbers for droplet formation between theory and experiment were attributed to an effective finite-range interaction through diffusion Monte-Carlo calculations~\cite{Cikojevic2020b}.

In low-dimensional systems, quantum fluctuations are enhanced leading to droplet formation and stabilization for wider parameter ranges, and independent of atom number 
\cite{Petrov2016,Astrakharchik2018,Chiquillo2018b,Chiquillo2019,Parisi2019,DeRosi2021,Mistakidis2021}. 
References~\cite{Zin2018,Ilg2018,Lavoine2021} addressed the effects of a cross-over to lower dimensions on the LHY-contributions.
The properties of the droplet phase eGP ground state (GS) and low-lying modes in one (or quasi-one) dimension (1D) were investigated in ~\cite{Cappellaro2018,Tylutki2020,NilssonTengstrand2022}, mapping out the collective modes across the homogeneous-to-droplet~\cite{NilssonTengstrand2022} and droplet-to-soliton transition~\cite{Cappellaro2018}.
Thermal instabilities have also been addressed~\cite{Wang2021,DeRosi2021}.
Corrections in 1D beyond LHY were discussed in Ref.~\cite{Ota2020a},  and alternative functionals based on 1D as well as 3D quantum Monte Carlo solutions were suggested~\cite{Cikojevic2020,Cikojevic2020b,Kopycinski2023}. 
Low dimensionality is generally favorable for exact approaches, as  exemplified long ago by the well known exact Lieb-Liniger model~\cite{LiebLiniger1963,Lieb1963} for a repulsive 1D single-component Bose gas.
Previous studies have applied quantum Monte Carlo techniques~\cite{Parisi2019,Parisi2020}, the Bose Hubbard model~\cite{Morera2020,Morera2021a}, and an effective quantum field theory~\cite{Chiquillo2018}. Additionally, the bosonic multi-configurational time-dependent Hartree approach has been applied to study droplet dynamics~\cite{Mistakidis2021}. 

Experimental realizations of self-bound boson droplet states have so far been restricted to systems of at least 
several dozen or hundreds of atoms. 
For fermionic systems, however, a new generation of micro-traps has enabled the realization of few-body  states~\cite{Serwane2011, Zurn2012, Zurn2013}. For such systems quantum tunneling processes~\cite{Rontani2012} allow 
 single-atom control, leading to observations of novel and strongly correlated few-body phases~\cite{Murmann2015} and making 
the study of complex many-body phenomena accessible to a bottom-up approach~\cite{Deuretzbacher2014,Wenz2013}. In theoretical work, excitations of such few-fermion systems~\cite{Bjerlin2016,Resare2022} have been interpreted as precursors of Higgs-Anderson (HA)-like amplitude modes~\cite{Higgs1964,Anderson1958} 
signaling the transition from a normal to a paired phase~\cite{Bruun2014,Bayha2020,Holten2022}. 
With this progress in mind, we here investigate the few- to many-body aspects of droplet formation in a
1D binary system of bosons. 
Applying an importance-truncated configuration interaction approach, we 
report numerically exact solutions for the ground state and low-lying excitations of a few-boson system, uncovering the  emergence of collective modes across criticality. 

This work is organized as follows. Section~\ref{sec:Model} defines the Hamiltonian on a ring for the bosonic system. The eGP approach is also briefly recalled.
Section~\ref{sec:Transition} provides the numerically exact few-body spectra and pair correlations for a system of in total eight bosons, and compares the few-body results with the eGP approach.  
Further evidence for the transition to a localized state is provided by the rotational spectra discussed in Sec.~\ref{sec:Yrast} and a discussion of dynamical properties in Sec.~\ref{sec:Dynamics}.
A summary is given in Sec.~\ref{sec:Summary} along with future perspectives. 

%%%%%%%%%%%%%%%%%%

\section{Model}\label{sec:Model}

We consider a binary bosonic mixture with components of equal mass $M$, such as the hyperfine states of $^{39}\mathrm{K}$, as in Refs.~\cite{Cabrera2018,Semeghini2018,Cheiney2018}. The components, labeled $\sigma \in \{A,B\}$, interact via the usual contact interactions with effective intra-species $(g_{AA}$, $g_{BB})$ and inter-species $(g_{AB})$ strength parameters. We impose the constraints of equal atom numbers ($N_A = N_B = N/2)$ 
and equal intra-species interactions $(g_{AA}=g_{BB}=g)$. 
Confining the system to a 1D ring of radius $R$, the Hamiltonian reads 
\begin{equation}\label{eqn:ExactHamiltonian}
    \begin{split}
        \hat{H} = &\sum_{\sigma,m} \frac{m^2}{2} \hat{a}_{\sigma,m}^\dagger \hat{a}_{\sigma,m} \\
        &+ \frac{1}{2}\sum_{\substack{\sigma,\sigma'\\m_1,m_2,k}}\frac{g_{\sigma\sigma'}}{2\pi} \hat{a}_{\sigma,m_1 +k}^\dagger \hat{a}_{\sigma',m_2 -k}^\dagger  \hat{a}_{\sigma',m_1}\hat{a}_{\sigma,m_2}.
    \end{split}
\end{equation}
Here setting $\hbar = M = R = 1$ defines the dimensionless units used throughout this work. The operators $\hat{a}_{\sigma , m}^\dagger (\hat{a}_{\sigma , m})$ create (annihilate) a boson of species $\sigma $ in the single-particle angular momentum eigenstate $\phi_{m} (\theta) = \frac{1}{\sqrt{2\pi}} e^{im\theta}$, where $\theta$ is the azimuthal position on the ring and $m$ is the integer one-body angular momentum quantum number.
In the limit of small $N$ direct diagonalization of the Hamiltonian Eq.~\eqref{eqn:ExactHamiltonian} becomes feasible, giving access to numerically exact solutions. 

Previous studies have used exact diagonalization methods to investigate the formation of solitons in a single-component attractive BEC for several hundred particles~\cite{kanamoto2003quantum,kanamoto2003stability,kanamoto2005symmetry}. In these cases the low lying excitation spectra could be captured by a one-body basis with only three \cite{kanamoto2003stability,kanamoto2005symmetry} or  five \cite{kanamoto2003quantum} single-particle states. 
{In the case considered here, {\it i.e.}, in the regime of droplet formation of a binary few-boson mixture described by Eq.~\eqref{eqn:ExactHamiltonian} we found that the considered states have comparatively large interaction energy contributions and are highly correlated, requiring the single-particle angular momentum cut-off to be large. We found ${|m| \leq m_\text{max} = 60}$ to be adequate for the interaction strengths and particle numbers considered in this work. }
The resulting one-body basis of size $121$
yields a Hilbert space that is prohibitively large for any naive direct diagonalization of the many-body Hamiltonian. This holds even when realizing its block-diagonal structure due to conserved total angular momentum $L$. 
A remedy to this problem is that for a given low-lying energy eigenstate within one of these blocks, only a relatively small subset of all many-body basis states has a non-negligible contribution
to the exact solution. The dimension of the relevant Hilbert space for a specific target eigenstate can thus be greatly reduced.
To identify the relevant elements of the Hilbert space for each desired energy state we employ a so called importance truncated configuration interaction (ITCI) method~\cite{Roth2009}. 
The exact Hamiltonian Eq.~\eqref{eqn:ExactHamiltonian} is diagonalized in a subspace of the Hilbert space to which 
many-body basis states are iteratively added. The condition to include a many-body basis state is 
that the magnitude of its overlap with a perturbative expansion of 
the considered energy eigenstate is greater than or equal to some predefined threshold. 
For the systems in this work, a threshold of $10^{-5}$ is considered adequate.
In this way the description is successively expanded, acquiring an increasingly accurate truncation of the Hilbert space tailored to the Hamiltonian~Eq.~\eqref{eqn:ExactHamiltonian} and to the desired energy eigenstate or small set of eigenstates. 
Details of the construction of the Hilbert spaces in which the exact diagonalization is performed along with all relevant convergence parameters can be found in Appendix~\ref{sec: Appendix ITCI}. {A list of selected ground-state and excitation energies is also provided.}

Where possible we draw comparisons between the exact and BMF results, with the aim of characterizing the few- to many-body transition in the homogeneous-to-droplet cross-over region.
For a binary system with components of equal atom numbers and equal intra-species interactions the eGP equation reads
\begin{equation}\label{eqn: order parameter diff eqn}
    \begin{split}
        \mu\Psi = &-\frac{1}{2}\frac{\partial^2\Psi}{\partial \theta^2} + \frac{N}{2}(g + g_{AB})\left|\Psi\right|^2 \Psi \\
        &- \frac{\sqrt{N}}{\pi 2^{3/2}}[(g + g_{AB})^{3/2} + (g - g_{AB})^{3/2}]\left|\Psi\right|\Psi
    \end{split}
\end{equation}
where $\mu $ is the chemical potential, and the normalization condition {for the order parameter} $\int|\Psi(\theta)|^2 d\theta = 1$ applies. 
A trivial solution to Eq.~\eqref{eqn: order parameter diff eqn} is the homogeneous solution $\Psi_0$, however this solution is not always stable. Specifically, energetic and dynamic instability occurs for $\Psi_0$ when
\begin{equation}\label{eqn:instability}
    \begin{split}
        0 > - \frac{\sqrt{N}}{8\pi^{3/2}}\Big[(g &+ g_{AB})^{3/2} + (g - g_{AB})^{3/2}\Big]\\
        & + \frac{N}{2\pi}\frac{(g + g_{AB})}{2} + \frac{1}{4},\\
    \end{split}
\end{equation}
defining the phase boundary between the homogeneous and localized phases. (See Appendix \ref{sec:Appendix} for a derivation of Eqs.~\eqref{eqn: order parameter diff eqn} and~\eqref{eqn:instability}.)

\section{Few-body transition from a homogeneous to a localized state}\label{sec:Transition}

\begin{figure}[h!]
    \centering
    \includegraphics[width=8.0cm]{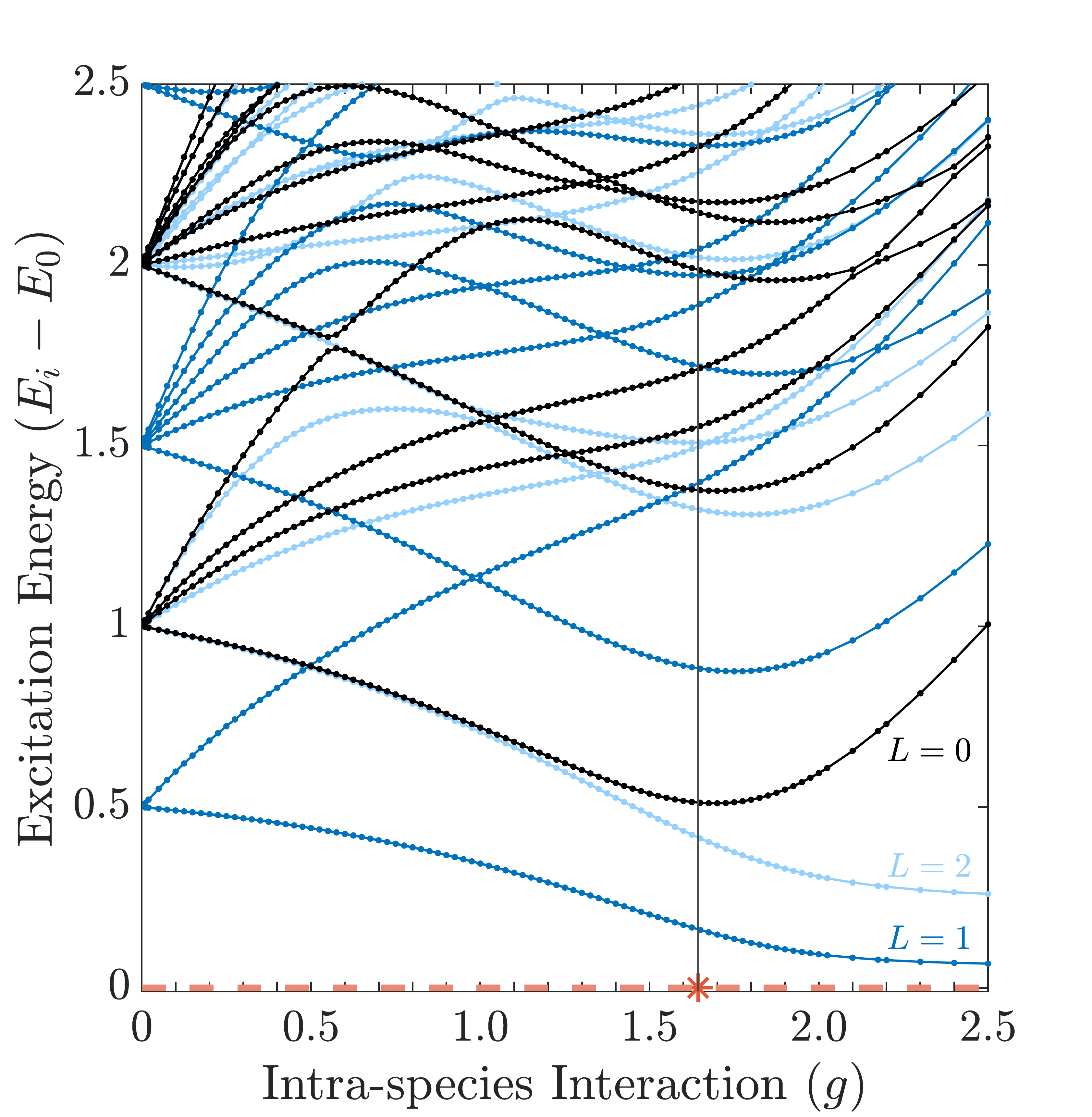}
    \caption{Low lying excitation energies, $E_i-E_0$, for states with total angular momentum $L = 0$ (black), $L = 1$ (blue), and $L = 2$ (light blue) for a system of $N = 8$ particles with $g_{AB} = -0.9g$ corresponding to the red dashed path through the phase diagram Fig.~\ref{fig:Phase}. 
    {The asterisk (with the black vertical line to guide the eye) marks the intersection of ${g_{AB} = -0.9}g$ with the BMF phase boundary}.
    The spectra contains degenerate energy levels (crossings) as well as avoided crossings. A more detailed analysis of the excited states is presented in Sec.~\ref{sec:Dynamics}.
The points that have been calculated are indicated by circular markers and the joining lines are
provided as a visual aid.
\bigskip
} 
%\end{figure}
%\begin{figure}[h!]
    \label{fig:SpectraA}
    \centering
    \includegraphics[width=8.0cm]{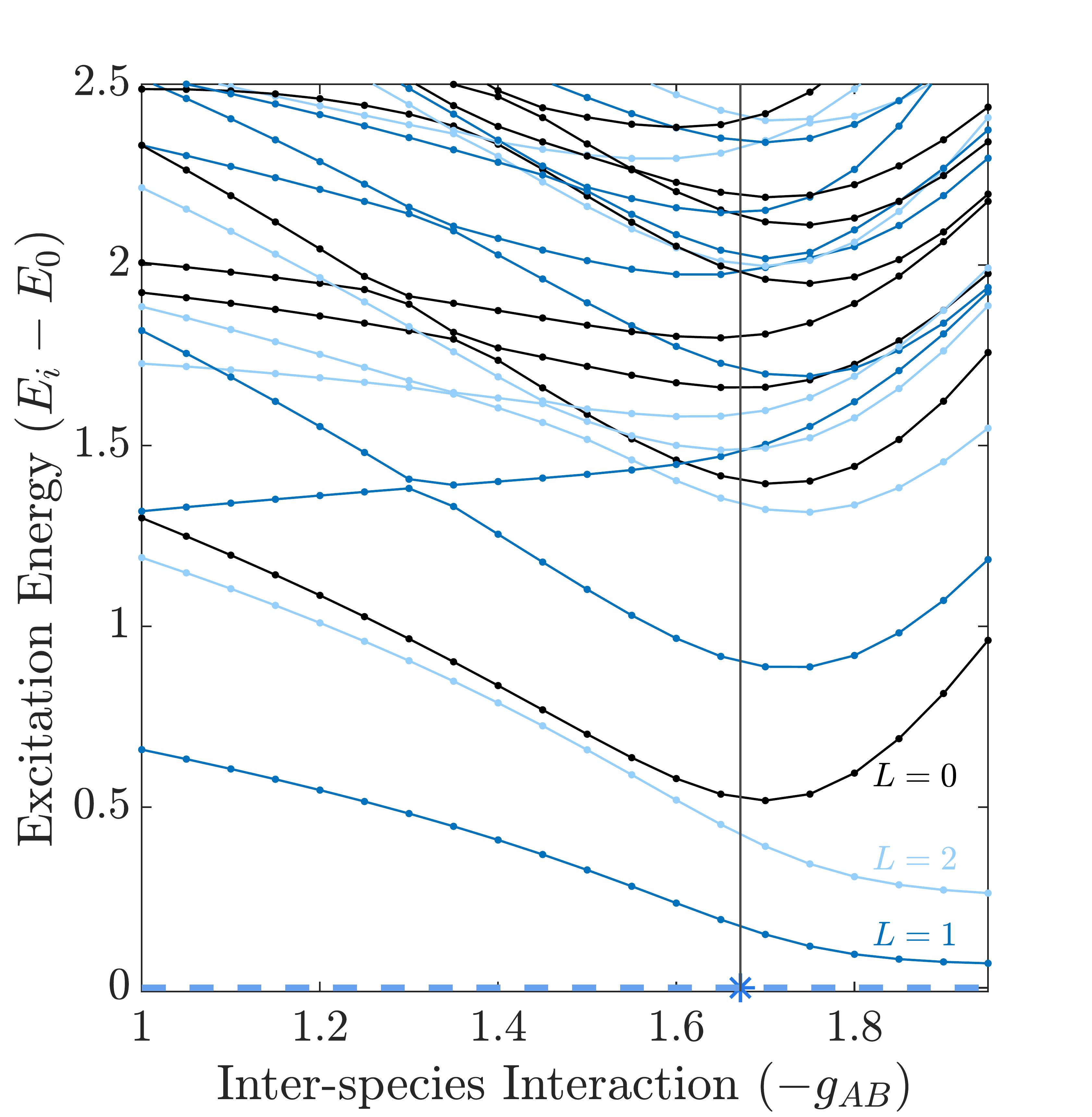}
    \caption{Excitation spectra as in Fig.~\ref{fig:SpectraA} but for fixed ${g=2}$ and variable $g_{AB}$, corresponding to the blue dashed path through the phase diagram of Fig.~\ref{fig:Phase}.}
    \label{fig:SpectraB}
\end{figure}

Let us now investigate the exact ground state and low-lying excitations of binary bosonic mixtures described by
Eq.~\eqref{eqn:ExactHamiltonian} in the  case of attractive inter-species interactions ${g_{AB} < 0}$ and equal
intra-species repulsion ${g = g_{AA} = g_{BB} > 0}$.
This choice of interaction strengths for the ITCI calculations is guided by BMF analysis 
(as detailed in Appendix \ref{sec:Appendix}).
The excitation energy spectra for a system of $N=8$ bosons with fixed relative attraction $g_{AB}/g=-0.9$ and variable intra-species repulsion ${0\le{g}\le 2.5}$ are shown in Fig.~\ref{fig:SpectraA}.
The spectra for the $N=8$ system with fixed $g=2$ and variable $g_{AB}$ are presented in Fig.~\ref{fig:SpectraB}.
In both cases the excitation energies are shown for the three lowest values of total angular momentum,  $L=0$ (black), $L=1$ (blue) and $L=2$ (light blue).
The excitation energies are taken relative to the ground state, which itself has total angular momentum $L=0$. 
All states were determined with the numerical accuracy of convergence specified in Appendix~\ref{sec: Appendix ITCI}.

A noticeable feature of both excitation spectra is the distinct minimum in the lowest $L=0$ excitation mode. Furthermore, the gaps between the ground state and the lowest rotational modes with $L=1$ and $L=2$, respectively, diminish upon increasing $g$ as in Fig.~\ref{fig:SpectraA}, or $-g_{AB}$, as in Fig.~\ref{fig:SpectraB} which will be discussed further in Sec.~\ref{sec:Yrast}. Interestingly, 
the low-lying excitations bear the typical signatures of a broken symmetry in the ground state when passing through a certain critical range of interaction strengths. 
In particular, the formation of a Higgs-Anderson (HA) -like gapped mode (such as the lowest $L=0$ mode of Fig.~\ref{fig:SpectraA}) has been seen in other cases of symmetry breaking, for example in soliton formation in a BEC~\cite{kanamoto2003quantum} or the BEC to supersolid phase transition~\cite{hertkorn2019fate}, as well as in the formation of paired fermions in the few- to many-body regime~\cite{Bjerlin2016,Bayha2020,Resare2022}. Here the formation of the broken symmetry state, manifesting as an excitation energy minimum, requires a sufficiently large interaction energy contribution (for fixed ratio $g_{AB}/g$), see Fig.~\ref{fig:SpectraA}. However, a large interaction energy contribution alone (\textit{e.g.,} $g=2$) is not sufficient, the inter-species attraction  must also be sufficiently large, see Fig.~\ref{fig:SpectraB}.

The few-body ground state solutions obtained by direct diagonalization of Eq.~\eqref{eqn:ExactHamiltonian} necessarily preserve the Hamiltonian's azimuthal symmetry. To further analyze the formation of a bound state hidden in the internal structure of the full eigenstate $|\Psi \rangle $ we must turn to correlation functions. 
By fixing the position, $\theta'$, of a single particle of species $\sigma'$ and calculating the probability distribution of all other particles of species $\sigma$ with respect to the fixed position, the pair correlations  
\begin{equation}\label{eqn: pair correlations}
    \begin{split}
        \rho^{(2)}_{\sigma\sigma'}(\theta,\theta') = \sum_{m,n,k,l}\phi_m^*&(\theta)\phi_n^*(\theta')\phi_{k}(\theta')\phi_{l}(\theta) \\
        &\times\langle \Psi| \hat{a}^\dagger_{\sigma,m}\hat{a}^\dagger_{\sigma',n}\hat{a}_{\sigma',k}\hat{a}_{\sigma,l}|\Psi\rangle
    \end{split}
\end{equation}
map out the internal structure of the quantum state.  We note that 
$\rho^{(2)}_\text{tot} = \rho^{(2)}_{AA} + \rho^{(2)}_{BA} = \rho^{(2)}_{BB} + \rho^{(2)}_{AB}$ for symmetric components $A$ and $B$ with the normalization condition $\int \rho^{(2)}_{\sigma\sigma'}(\theta,\theta') d\theta = {N_\sigma-\delta_{\sigma\sigma'}}$.

\begin{figure}[h!]
    \centering
    \includegraphics[width=8.5cm]{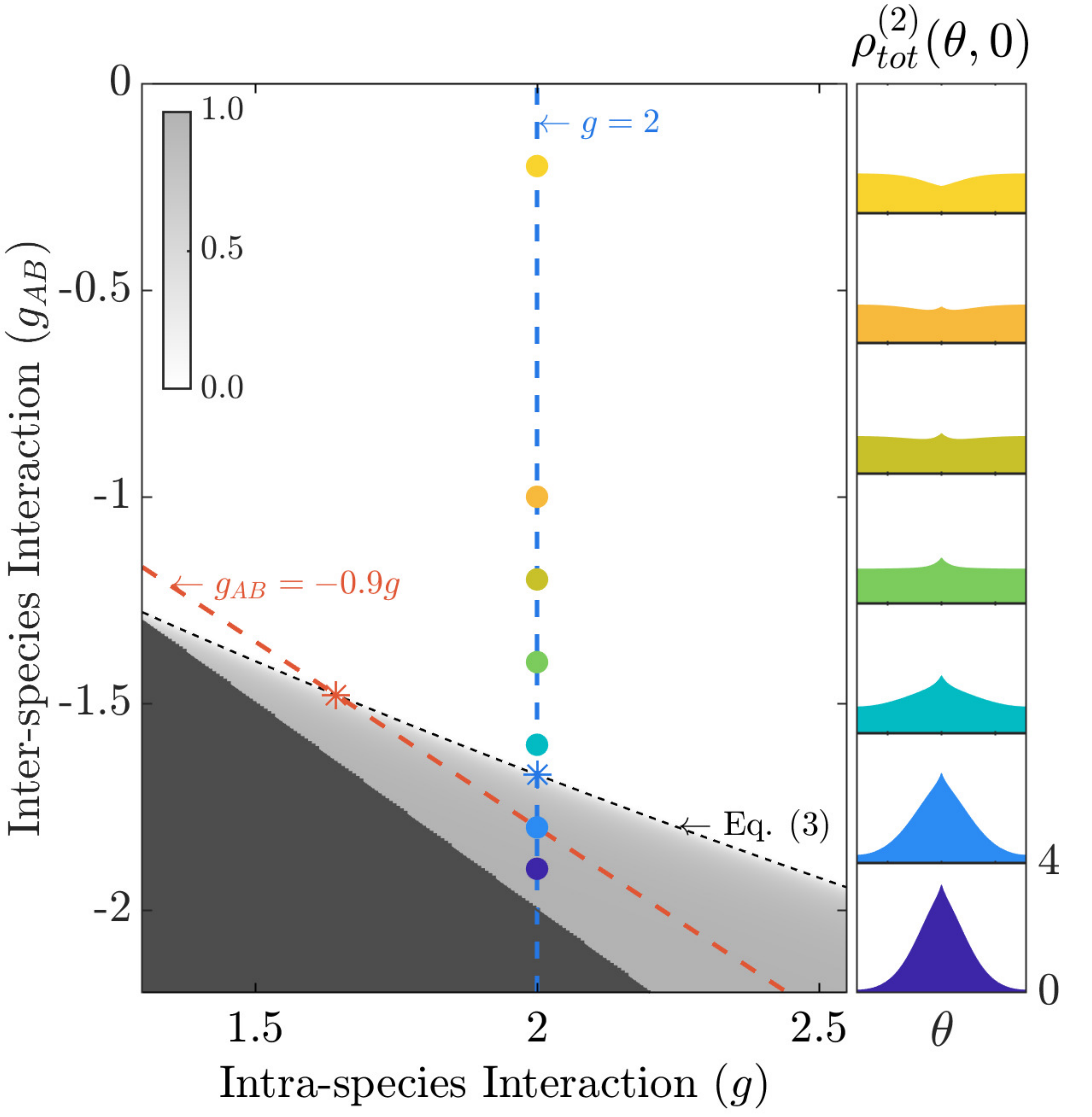}
    \caption{Transition from homogeneous to droplet states for $N = 8$. Sets of interaction parameters are represented by colored dots, and the corresponding pair correlations for the ground state with zero angular momentum are shown in the right column. The reference particle's position is $\theta' = 0$. The  phase  boundary Eq.~\eqref{eqn:instability}
    is plotted as a dashed black curve in the left diagram. The red line ${(g_{AB} = -0.9g)}$ identifies the path of parameters chosen in Fig.~\ref{fig:SpectraA} and the blue line ${(g = 2)}$ identifies the path of parameters considered in Fig.~\ref{fig:SpectraB}. The asterisks mark their intersection with the BMF phase boundary.
    The background indicates the contrast $(\frac{n_\text{max} - n_\text{min}}{n_\text{max} + n_\text{min}})$ of the eGP ground state at each point in the phase diagram. $n_\text{max}$ and $n_\text{min}$ are the maximum and minimum values of the eGP ground state density respectively. (The black area, $g-g_{AB}<0$, is outside the region of validity of the eGPE, Eq.~\eqref{eqn: order parameter diff eqn}.)}
    \label{fig:Phase}
\end{figure}

Let us now investigate the  $L=0$ ground state.
Figure~\ref{fig:Phase} (left panel) shows the phase diagram for a system of $N=8$ bosons. The colored dots along the vertical line $g=2$ indicate representative values of $g_{AB}$. For each of these sets of interaction parameters the corresponding ground state pair correlations are shown in the same color in the right panel.  
Beginning with the weakly attractive inter-species interaction marked by the yellow dot we see that the associated pair correlation for the ground state (also in yellow)
has a small indent at the location of the fixed particle $(\theta' = 0)$. Here, the inter-species attraction cannot compensate for the repulsion of the atoms of the same species. As the strength of the inter-species attraction increases a peak begins to form at the position of the fixed particle. The peak becomes more pronounced with increasing inter-species attraction (see pair correlations plotted in green and light-blue) showing the onset of localization. Beyond $g_{AB}\lesssim  -1.6$ the pair correlations indicate  the formation of a localized state (blue and dark-blue plots).
While these are exact few-body results, an analogy can be drawn to the quantum liquid droplets previously observed for large particle numbers~\cite{Cabrera2018,Semeghini2018}. In one dimension such droplets form with a repulsive mean-field interaction and are stabilized by the next-order quantum correction, the LHY term, which in this case is attractive~\cite{Petrov2015,Petrov2016}.
The phase boundary Eq.~(\ref{eqn:instability}) derived from the condition for energetic and dynamic stability of the homogeneous solution to the eGP equation (see Appendix \ref{sec:Appendix}) is plotted as a dashed black curve in Fig.~\ref{fig:Phase}. 
Below this curve the homogeneous solution is unstable.
This is the region of droplet formation predicted by the BMF theory.

To further illustrate the presence of a phase transition we plot the 
{ contrast $({n_\text{max} - n_\text{min}})/({n_\text{max} + n_\text{min}})$ }
as calculated from the maximum and minimum density, $n_\text{max}$ and $n_\text{min}$, 
of the numerical ground state solution of the eGP equation at each point in the phase diagram.
The consistency between the ground state pair correlations of the exact results and the phase transition predicted by the BMF theory is noteworthy.
The red and blue dashed lines denote $g_{AB}= -0.9g$ and $g = 2.0$, respectively. The intersections of the BMF phase boundary with these paths through the phase diagram are marked by asterisks. These are the points of phase transition along the respective paths as predicted by the BMF model. The associated values of $g$ and $g_{AB}$ are plotted as vertical black lines in 
Fig.~\ref{fig:SpectraA} and Fig.~\ref{fig:SpectraB} respectively for comparison with the exact low-lying energy modes.
In both spectra the minimum of the lowest zero angular momentum mode approximately coincides with the BMF transition point.
In fact these minima 
signal the few-body precursor of a phase transition in the low energy spectra.
We note that the broad shallow minima seen in Fig.~\ref{fig:SpectraA} and Fig.~\ref{fig:SpectraB} are indicative of the ambiguity of the precise point of criticality in the few-body limit.
\begin{figure}[h!]
    \centering
    \includegraphics[width=8cm]{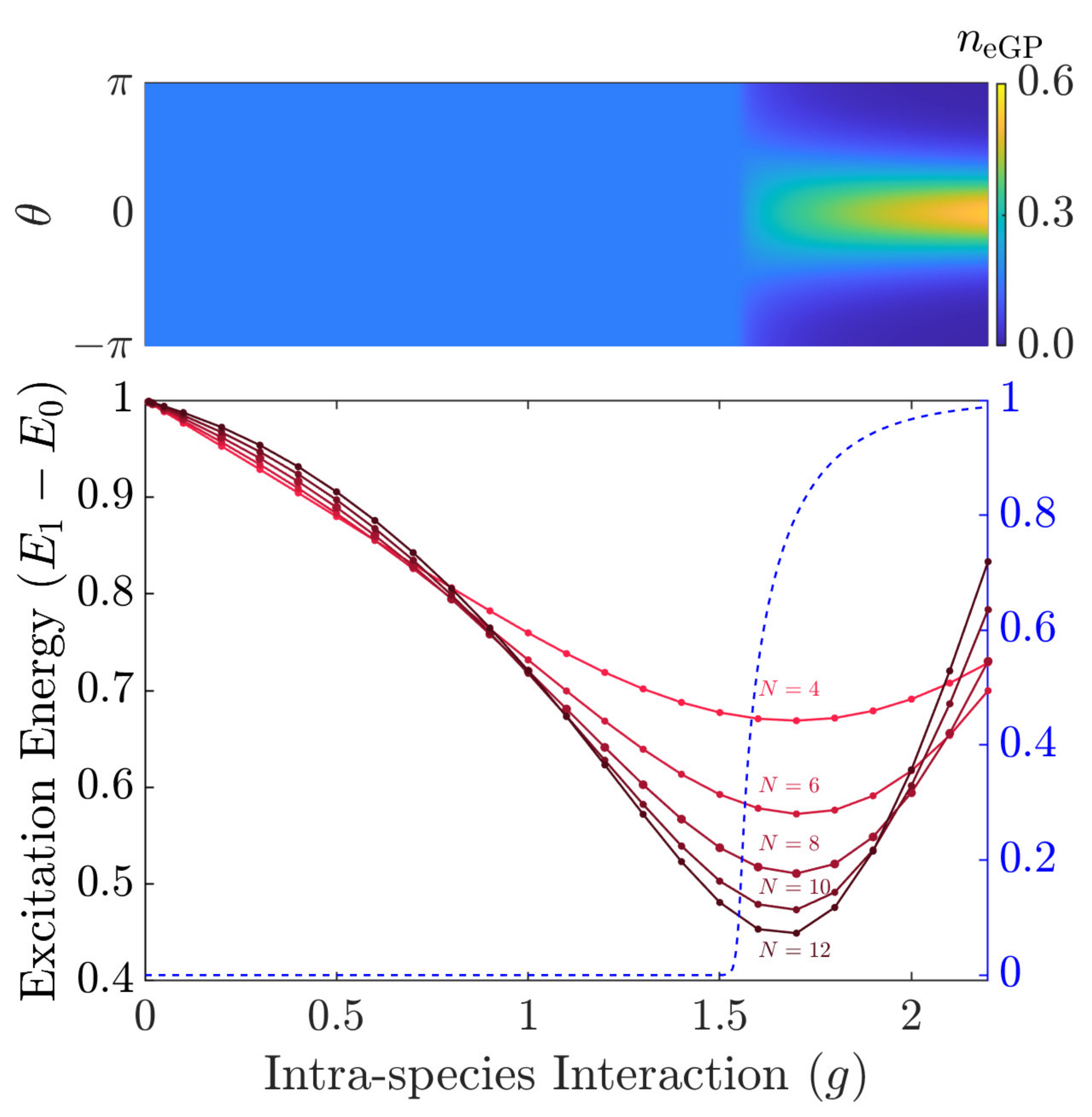}
    \caption{ \textit{Upper panel:} Contour plot of the density $n_{\mathrm{eGP}}$ of the numerical ground state solution of the eGP equation with $N=12$ as a function of $g$.
    \textit{Lower panel:} { The excitation energies of the lowest $L=0$ mode  obtained with the ITCI approach for ${4 \le N\le 12}$ (left axis) and the contrast of the eGP ground state shown in the upper panel (right axis) for $N=12$.} }
    \label{fig:Higgs}
\end{figure}

The lowest zero angular momentum mode for various $N$ are shown in the lower panel of Fig.~\ref{fig:Higgs},  
together with the contrast of the eGP ground state (dashed line).
The eGP ground state density $n_{\mathrm{eGP}}$ is shown as a function of $g$ for parameters corresponding to $N=12$ in the upper panel.
With increasing $N$ the minimum of the exact few-body mode deepens, accompanied by a shift of the critical value to slightly smaller $g$, approaching the point of transition in the BMF limit. 

\section{Rotational Spectra}\label{sec:Yrast}
 
 We saw above how the internal structure of the exact ground state not only manifests in the pair correlations but is also reflected in the signatures of symmetry breaking seen in the excitations of the system.
An internally-broken spatial symmetry, of which the localization of particles into a bound bosonic state on a ring is a particularly clear example, will have a prominent effect on the rotational excitations and the energy dispersion as a function of angular momentum (the so-called ``yrast" line), as is well known  from nuclear structure theory~\cite{BohrMottelson}. 
For the ring system studied here, total angular momentum is conserved and we proceed to analyze the low-lying excitation energies as functions of $L$.
Due to the periodic boundary conditions the system satisfies Bloch's theorem~\cite{Bloch1973}. The energy spectra can thus be expressed as the sum of a parabolic term $L^2/(2NMR^2)$, corresponding to the kinetic energy of a rigid body of mass $NM$ rotating around the circumference of the ring, and a term that is periodic in $L$ with 
periodicity $L = N$~\cite{Bloch1973,Smyrnakis2009}. 
\begin{figure}[h!]
    \centering
    \includegraphics[width=9.2cm]{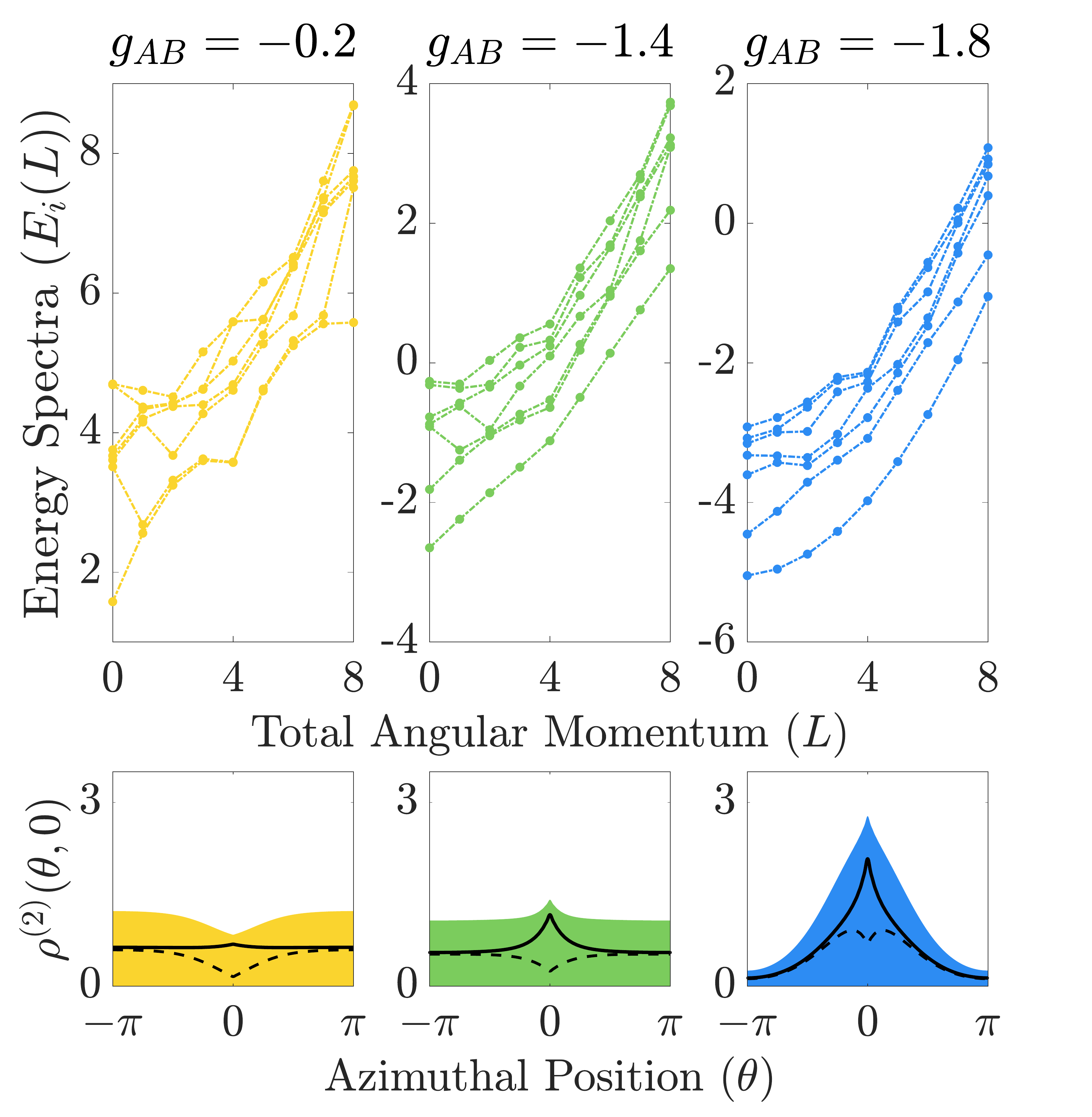}
    \caption{{\it Upper panel:} Exact energy spectra (ground state and six lowest excitations) 
  as a function of total angular momentum $(L)$ for a system of $N = 8$ particles with $g = 2$ and $g_{AB} = -0.2, -1.4, -1.8$.  {\it Lower panel:} Pair correlations for $L = 0$ ground states. The total pair correlations $\rho^{(2)}_\text{tot}$ are plotted in  color, corresponding to the scheme introduced in Fig.~\ref{fig:Phase}. The pair correlations for the species of the fixed component atom, 
  $\rho^{(2)}_{AA}$, are presented as dashed lines. The pair correlations of the opposite species, $\rho^{(2)}_{BA}$, are presented as solid lines. }
    \label{fig:YrastSpectra}
\end{figure}
Furthermore, the periodic component of the energy is symmetric about $L = N/2$ which is a result of the invariance of the two-body interaction term of Eq.~\eqref{eqn:ExactHamiltonian} under the transformation $m \rightarrow 1-m$ of all one-body angular momentum quantum numbers. This transformation maps many-body basis states with total angular momentum $L = \sum_n m_n$ and energy $E$ to states with total angular momentum $L' = N - L$ and energy $E' = E - L + N/2$~\cite{Smyrnakis2009}. 
The upper panel in Fig.~\ref{fig:YrastSpectra} shows the total energy spectra for one period in angular momentum for $g=2$ and $g_{AB}=-0.2$, $-1.4$ and $-1.8$, corresponding to the yellow, green, and blue points in the phase diagram of Fig.~\ref{fig:Phase}. The lower panel shows the pair correlations for the $L=0$ ground states with $\rho^{(2)}_\text{tot}$ in color, $\rho^{(2)}_{AA}$ in dashed black and $\rho^{(2)}_{BA}$ in solid black lines.

A hallmark of superfluidity on a ring is persistent dissipation-less flow. These  states occur 
due to local minima in the ground state energy at finite angular momenta~\cite{Bloch1973,Anoshkin2013}. In the top left panel of Fig.~\ref{fig:YrastSpectra} indeed one observes a very shallow local minimum in the yrast line at total angular momentum $L = 4$. 
This negative yrast line curvature 
is indicative of a few-body precursor of states supporting persistent currents.
From the three plots in the upper panel of Fig.~\ref{fig:YrastSpectra} one can clearly see that increasing the inter-species attraction for fixed intra-species repulsion drives a change in the yrast line curvature, from  superfluid-like (left) to an intermediate regime (center) and finally to a nearly parabolic yrast line (right) that is indicative of rigid-body rotation.
(Such changes of the yrast line have also been discussed in the context of supersolidity in toroidally trapped dipolar 
condensates~\cite{NilssonTengstrand2021}, asymmetric ring condensates~\cite{Ogren2021}, ring-trapped 
droplet-superfluid compounds~\cite{NilssonTengstrand2022,Holmstrom2022} and mixed bubbles in bosonic mixtures~\cite{Sturmer2022}.)
The ground state pair correlations shown in the lower panel for all three cases confirm the transition from a 
homogeneous to a localized state.

Let us now revisit Fig.~\ref{fig:SpectraA} in which we have plotted the excitation energies  in the range ${0\leq g \leq 2.5}$ for fixed ${g_{AB}/g=-0.9}$, 
{\it i.e.}, along the red path through the phase diagram of Fig.~\ref{fig:Phase}. 
As $g$ increases into the droplet regime, the lowest energy modes for total angular momentum $L\neq 0$ flatten and approach some constant energies, corresponding to the rigid body rotation of the droplet indicated by the parabolic yrast-line curvature. 
If one could neglect the kinetic energy cost of rotating the localized state around the circumference of the ring, these modes would become degenerate ground states. 
Indeed, the rotational modes scale as $\Delta E= {L^2}/{(2NMR^2)}$ and may  be considered massless in the thermodynamic limit. We therefore argue that the lowest rotational modes of Fig.~\ref{fig:SpectraA} may be interpreted as  few-body precursors of Goldstone modes~\cite{Goldstone1961}.
Remarkably, the region of steepest slope of these  rotational modes and the minimum of the lowest zero angular momentum mode approximately coincide with the BMF prediction of the transition point.

\section{Dynamical properties}
\label{sec:Dynamics}

We proceed to study the system dynamics in response to modulations in the transition-driving parameters, {\it i.e.},  $g_{AA}$, $g_{BB}$ and $g_{AB}$. Such dynamics can be crucial for experimentally  observing
excitation modes, as recently bared out in Ref.~\cite{Bayha2020} 
in the context of a few-fermion system. 
In this section we restrict our analysis to interaction driven excitations from the many-body ground state
with $g_{AB} = -0.9g$ in the $L=0$ subspace. A complete understanding of these dynamics requires a detailed analysis of the ground state and low lying energy eigenstates, in particular their behavior under the transformations which define three global symmetries of the system: interchange of distinguishable species, reflection of all one-body angular momenta \textit{i.e.,} $m\rightarrow -m$, and the continuous rotational symmetry of the ring.
Let us consider these transformations in more detail.
Since we consider equal intra-species interactions ${(g_{AA} = g_{BB} = g)}$ and equal atom numbers ${(N_A = N_B)}$ the system has a global symmetry corresponding to the interchange of the two distinguishable species. We may exchange all atoms of component $A$ for atoms of component $B$ and all atoms of component $B$ for atoms of component $A$ without altering {the}  physical properties of the system. Two such exchanges of the species labels must return any state to its original form. Therefore all non-degenerate energy eigenstates are either symmetric or anti-symmetric with respect to the interchange of the {distinguishable} species. 
The second global symmetry of the system is the reflection of all one-body angular momentum quantum numbers ${m\rightarrow -m}$. Clearly two applications of the transformation returns any state to its original form. This transformation maps energy eigenstates with total angular momentum $L\neq 0$ to degenerate energy eigenstates with total angular momentum $-L$.
However, for eigenstates with $L=0$ no such degeneracy is guaranteed. Therefore all non-degenerate energy eigenstates with $L=0$ must be either symmetric or anti-symmetric with respect to reflection of all one-body angular momenta ($m$-reflection). 
In addition to these two symmetries, the system has the continuous rotational symmetry of the ring which ensures that every energy eigenstate has an integer total angular momentum. A modulation of the interactions strengths ($g_{AA}, g_{BB}, g_{AB}$, or any combination of these) preserves the total angular momentum $L$. Hence, systems in the many-body ground state subject to a modulation of the interaction parameters will remain in the $L=0$ subspace.
In the spectra presented in Figs.~\ref{fig:SpectraA} and~\ref{fig:SpectraB} avoided energy crossings are seen only between states with the same behavior with respect to all three symmetries.

To illustrate these distinct symmetries we consider the low-lying excitations in the limit of perturbatively week interactions $g = \delta g \gtrsim 0$.
In the non-interacting case, when ${g = g_{AB} = 0}$, the homogeneous many-body ground state is simply  ${|\Psi_0^{(g=0)} \rangle= |0^4\rangle_A|0^4\rangle_B}$ with energy $E_0 = 0$. The superscripts here denote the occupation of four bosons in the orbital $m=0$ for each of the species $A$ and $B$, respectively. We see immediately that ${|\Psi_0^{(g=0)} \rangle}$ is symmetric with respect to both species interchange and $m$-reflection. Furthermore, in the absence of any ground-state crossings (see Figs.~\ref{fig:SpectraA} and~\ref{fig:SpectraB}) these symmetry properties of the ground-state persist for all considered values of $g$.
The excitations in the non-interacting case are straightforwardly described in terms of momentum-conserving particle excitations.
In particular, the first $L=0$ excited state has energy $E_1 = 1$ and is four-fold degenerate, with a space spanned \textit{e.g.,} by 
\begin{equation}
    \begin{split}
        |\Psi_{1_a}^{(g=0)}\rangle &= |{-1}^1,0^2,{1}^1\rangle_A |0^4\rangle_B,\\
        |\Psi_{1_b}^{(g=0)}\rangle &= |0^4\rangle_A |{-1}^1,0^2,{1}^1\rangle_B,\\
        |\Psi_{1_c}^{(g=0)}\rangle &= |{-1}^1,0^3\rangle_A |0^3,{1}^1\rangle_B,\\
        |\Psi_{1_d}^{(g=0)}\rangle &= |0^3,{1}^1\rangle_A |{-1}^1,0^3\rangle_B.\\
    \end{split}
    \label{eqn: non-interacting excitations}
\end{equation}
We see that a transition from {\it e.g.}, $|\Psi_0^{(g=0)}\rangle$ to $|\Psi_{_{1_a}}^{(g=0)}\rangle$ is accessed via the intra-species interaction of species $A$ with interaction parameter $g_{AA}$.

For perturbatively weak interaction strengths 
${(g = \delta g \gtrsim 0)}$
the degeneracy of the first excitation is lifted and,
by numerical diagonalization of the Hamiltonian Eq.~\eqref{eqn:ExactHamiltonian} with ${g_{AA} = g_{BB} = \delta g \gtrsim0}$ and ${g_{AB}/\delta g = -0.9}$, we find
states with well-defined symmetries,
\begin{align}
    \begin{split}\label{eqn: psi0 weakly interacting limit}
        |\Psi_0^{(g = \delta g)}\rangle \approx |\Psi_{0}^{(g=0)}\rangle, \\
    \end{split}\\
    \begin{split}\label{eqn: HA weakly interacting limit}
        |\Psi_1^{(g = \delta g)}\rangle \approx \frac{1}{2}&[|\Psi_{1_a}^{(g=0)}\rangle + |\Psi_{1_b}^{(g=0)}\rangle \\
        &+ |\Psi_{1_c}^{(g=0)}\rangle + |\Psi_{1_d}^{(g=0)}\rangle ],\\
    \end{split}\\
    \begin{split}\label{eqn: psi2 weakly interacting limit}
        |\Psi_2^{(g = \delta g)}\rangle \approx \frac{1}{\sqrt{2}}[|\Psi_{1_a}^{(g=0)}\rangle - |\Psi_{1_b}^{(g=0)}\rangle],\\
    \end{split}\\
    \begin{split}\label{eqn: psi3 weakly interacting limit}
        |\Psi_3^{(g = \delta g)}\rangle \approx \frac{1}{\sqrt{2}}[|\Psi_{1_c}^{(g=0)}\rangle - |\Psi_{1_d}^{(g=0)}\rangle],\\
    \end{split}\\
    \begin{split}\label{eqn: psi 4 weakly interacting limit}
        |\Psi_4^{(g = \delta g)}\rangle \approx \frac{1}{2}&[|\Psi_{1_a}^{(g=0)}\rangle + |\Psi_{1_b}^{(g=0)}\rangle \\
        &- |\Psi_{1_c}^{(g=0)}\rangle - |\Psi_{1_d}^{(g=0)}\rangle ].\\
    \end{split}
\end{align}
In particular, 
$|\Psi_j^{(g = \delta g)}\rangle$ is symmetric with respect to species interchange for ${j=0,1,4}$ and anti-symmetric for ${j=2,3}$. Additionally, $|\Psi_j^{(g = \delta g)}\rangle$ is symmetric with respect $m$-reflection for ${j=0,1,2,4}$ and anti-symmetric for ${j=3}$.
(Equations~\eqref{eqn: HA weakly interacting limit}-\eqref{eqn: psi 4 weakly interacting limit} equally span the space of degenerate first excited states. We choose the basis Eq.~\eqref{eqn: non-interacting excitations} to explicitly illustrate the symmetries in Eqs.~\eqref{eqn: psi0 weakly interacting limit}-~\eqref{eqn: psi 4 weakly interacting limit}.)
In the higher excitations one finds all four possible combinations of these symmetries. For example, $|\Psi_{10}^{(g = \delta g)}\rangle$ is symmetric with respect to species interchange and anti-symmetric with respect to reflection of all 
one-body angular momenta.

Equations~\eqref{eqn: HA weakly interacting limit}-\eqref{eqn: psi 4 weakly interacting limit} are connected to excitations in the droplet phase via an adiabatic increase of $g$ and $g_{AB}$. In particular, $|\Psi_{1}^{(g = \delta g)}\rangle$ in Eq.~\eqref{eqn: HA weakly interacting limit} is the weakly interacting limit of the Higgs-Andersson (HA)-like mode. Once again, in the absence of any mode crossings the symmetry properties of  $|\Psi_{1}^{(g = \delta g)}\rangle$ persist for all considered values of $g$. Therefore the HA-like mode is symmetric with respect to both species interchange and reflection of all one-body angular momenta.
\begin{figure}[h!]
    \centering
    \includegraphics[width=8cm, trim={0.8cm 0 17cm 0},clip]{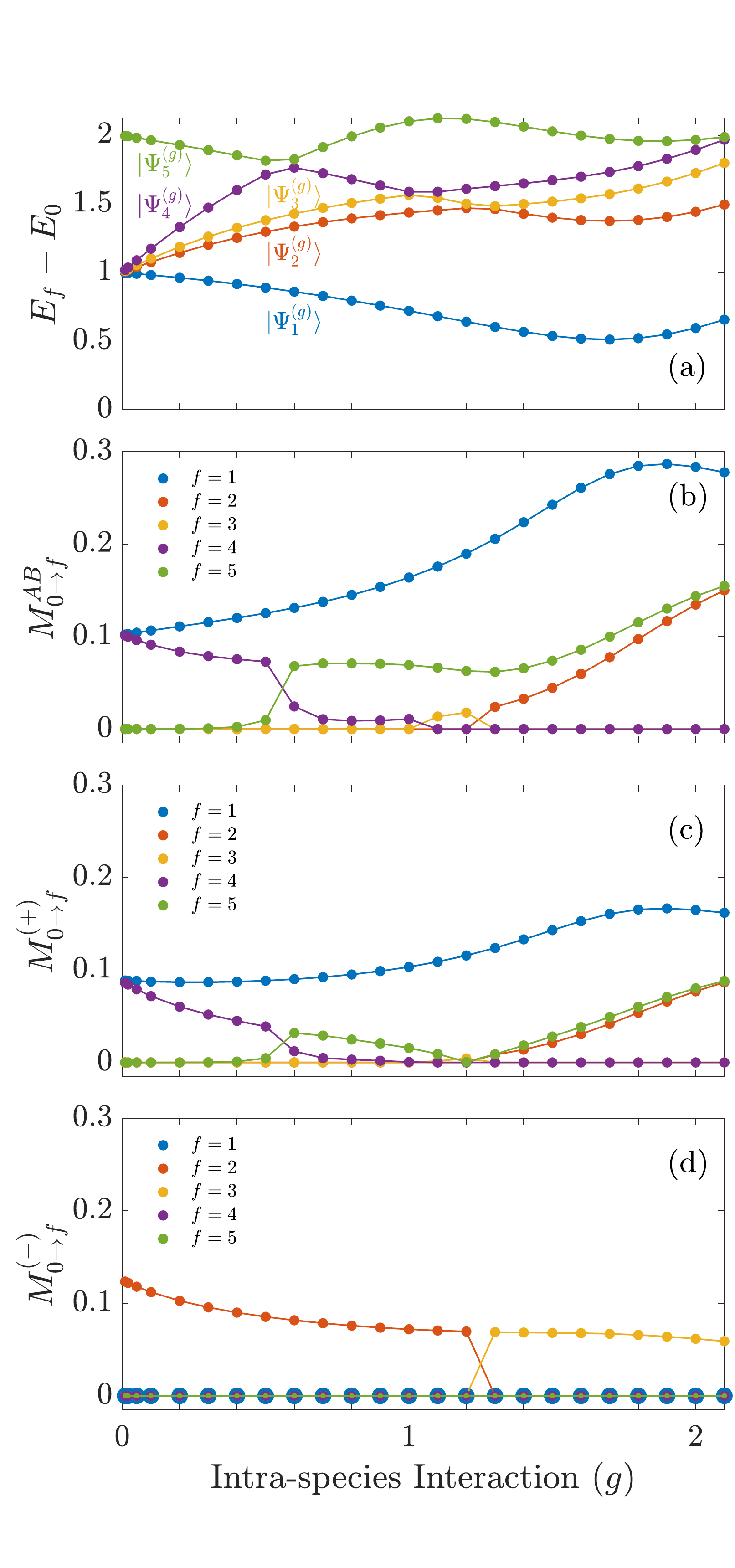}
    \caption{(a) Excitation energies ${E_f - E_0}$ and the associated transition matrix elements $M_{0\rightarrow f}^{AB}$ (b), $M_{0\rightarrow f}^{(+)} = M_{0\rightarrow f}^{AA} + M_{0\rightarrow f}^{BB}$ (c) and $M_{0\rightarrow f}^{(-)} = M_{0\rightarrow f}^{AA} - M_{0\rightarrow f}^{BB}$ (d) from the ground state to excitation $f$ with $L=0$ for various intra-species interaction strengths $g$ and fixed $g_{AB}/g = -0.9$. The points that have been calculated are indicated by circular markers and the joining lines are provided as a visual aid. Each excitation is plotted in a different color so that it may be associated with the relevant transition matrix elements. (d) ${M_{0\rightarrow f}^{(-)} =0}$ for ${f=1,4,5}$ and are therefore plotted in various sizes for visibility.}
    \label{fig: Matrix Elements}
\end{figure}
$|\Psi_{2}^{(g = \delta g)}\rangle$ in Eq.~\eqref{eqn: psi2 weakly interacting limit} is the weakly interacting limit of the third excitation in the droplet phase due to a mode crossing at $g\approx 1.3$, 
as can be seen in Fig.~\ref{fig:SpectraA} and  in Fig.~\ref{fig: Matrix Elements}(a). 
Therefore $|\Psi_3^{(g \gtrsim 1.3)}\rangle$ has the symmetry properties of Eq.~\eqref{eqn: psi2 weakly interacting limit}. Namely, it is anti-symmetric with respect to species interchange and symmetric with respect to $m$-reflection.
We shall see that the 
symmetry properties of Eqs.~\eqref{eqn: HA weakly interacting limit} and~\eqref{eqn: psi2 weakly interacting limit} 
reflect the breathing mode dynamics of their associated states in the droplet phase in superposition with the ground state.

First, we determine which states may be populated from the ground state via periodic modulations of the interaction parameters $g_{AA}$, $g_{BB}$ and $g_{AB}$. 
For weak modulation amplitudes we may use first order perturbation theory.
The transition rates are then obtained from the transition matrix elements and Fermi's golden rule. 
In particular, for a system that is initially $(t=0)$ in the  ground state $|\Psi _0^{(g)}\rangle$ and then at $t>0$ is subject to a periodic modulation of the interaction strength
\begin{equation}
    g_{\sigma\sigma'}(t) = g_{\sigma\sigma'}(0) + \eta \sin( \omega t),
    \label{eqn: modulation of gAB}
\end{equation}
where $\eta$ is a small amplitude, Fermi's golden rule gives the transition rate
\begin{equation}
    \begin{split}
        R^{\sigma\sigma'}_{0 \to f} \propto \eta^2\bigl (M_{0\rightarrow f}^{\sigma\sigma'})^2\times\delta(E_f - E_0 - \omega),\label{eqn: R^AB}
    \end{split}
\end{equation}
from $|\Psi_0^{(g)}\rangle$ to the state $|\Psi_f^{(g)}\rangle$.

In Fig.~\ref{fig: Matrix Elements} we present the transition matrix elements from the ground state to the $f^{\mathrm {th}}$ excited state with $L=0$, 
\begin{equation}
    M_{0\rightarrow f}^{\sigma\sigma'}  = |\langle \Psi_f^{(g)} | I_{\sigma\sigma'}|\Psi_0^{(g)} \rangle|, 
    \label{eqn: transition matrix elements}
\end{equation}
for the five lowest excitations at various values of $g$. Here
\begin{align}
    \begin{split}\label{eqn: I sigma sigma}
        I_{\sigma\sigma} = \sum_{i>j}\delta(\theta_{\sigma,i} - \theta_{\sigma, j}),
    \end{split}\\
    \begin{split}\label{eqn: I sigma sigmaBar}
        I_{\sigma\Bar{\sigma}} = \sum_{i,j} \delta(\theta_{\sigma,i} - \theta_{\Bar{\sigma},j}),
    \end{split}
\end{align}
are the operators associated with the two-body interactions and $\Bar{\sigma}$ refers to the opposite species of $\sigma$. 

In Fig.~\ref{fig: Matrix Elements}(a) we show the five lowest ${L=0}$ excitations from the spectra presented in Fig.~\ref{fig:SpectraA}, where ${g_{AB} = -0.9g}$. Here each excitation is plotted in a different color so that it may be associated with the relevant transition matrix elements.
In Figs.~\ref{fig: Matrix Elements}(b)-(d) we show the transition matrix elements $M^{AB}_{0\rightarrow f}$, ${M^{(+)}_{0 \rightarrow f} =M^{AA}_{0\rightarrow f} + M^{BB}_{0\rightarrow f}}$ and ${M^{(-)}_{0 \rightarrow f} = M^{AA}_{0\rightarrow f} - M^{AB}_{0\rightarrow f}}$, respectively. These matrix elements may be understood in terms of the system's symmetries. Beginning with the inter-species interaction ${(\sigma=A, \sigma'=B)}$, the two-body interaction operator $I_{AB}$ of Eq.~\eqref{eqn: I sigma sigmaBar} preserves both the $m$-reflection and the species interchange symmetries of the state it acts upon (\textit{e.g.,} $|\Psi_0^{(g)}\rangle$ as in Eq.~\eqref{eqn: transition matrix elements}). Hence modulation of $g_{AB}$, as in Eq.~\eqref{eqn: modulation of gAB}, can only induce transitions from the ground state to states with the same symmetry properties {\it i.e.}, states that are symmetric with respect to both global symmetries. This is in agreement with the computed transition matrix elements. At low $g$ only the states $|\Psi_1^{(g = \delta g)}\rangle$ and $|\Psi_4^{(g = \delta g) }\rangle$ may be accessed from the ground state via a periodic modulation of $g_{AB}$.  

We next consider excitations of the system by a periodic modulation of the intra-species interactions. Here, we distinguish between the two cases
of in-phase (``$+$") and out-of-phase  (``$-$") modulation, 
\begin{equation}
    \begin{split}
        g_{AA}(t) = g(0) + \eta \sin(\omega t),\\
        g_{BB}(t) = g(0) \pm \eta \sin(\omega t),\label{gbbt}
    \end{split}
\end{equation}
with associated operators ${I^{(\pm)} = I_{AA} \pm I_{BB}}$ and transition rates $R^{(\pm)}_{0 \to f}$ analogous to Eq.~\eqref{eqn: R^AB}. Again $\eta$ is a small amplitude. 
We first note that $I_{\sigma\sigma}$, $I_{\sigma\Bar{\sigma}}$ and all linear combinations of these operators preserve the $m$-reflection symmetry of the states they act on. Therefore any states that are anti-symmetric with respect to reflection of all one-body angular momenta are inaccessible from the ground state via any modulation of the interaction parameters corresponding to some linear combination of $I_{AA}$, $I_{BB}$, and $I_{AB}$.
This is why all computed transition matrix elements from $|\Psi_0^{(g = \delta g)}\rangle$ to $|\Psi_3^{(g = \delta g)}\rangle$ are strictly zero (see Fig.~\ref{fig: Matrix Elements}((b)-(d)).
While $I_{AA}$ and $I_{BB}$ preserve the $m$-reflection symmetry, both interactions break the species interchange symmetry, allowing access to different excitations via different combinations of these operators.
Furthermore, ${\langle\Psi_j^{(g)}|I_{AA}|\Psi_i^{(g)}\rangle = \pm \langle\Psi_j^{(g)}|I_{BB}|\Psi_i^{(g)}\rangle}$ where $``+"$ holds for states $i,j$ of the same species interchange symmetry and $``-"$ holds for states $i,j$ of opposite symmetries.
Thus the linear combination ${I^{(+)} = I_{AA} + I_{BB}}$ associated with in-phase modulation is symmetry preserving with respect to both $m$-reflection and species interchange.
Therefore by in-phase modulation of $g_{AA}$ and $g_{BB}$ we may access the same states as with the periodic modulation of $g_{AB}$. This is barred out by the comparison of Figs.~\ref{fig: Matrix Elements}(b) and (c).
In contrast, the operator ${I^{(-)} = I_{AA} - I_{BB}}$ associated with out-of-phase modulation maps states that are symmetric with respect to species interchange to states that are anti-symmetric and vice versa. Therefore, only states that are symmetric with respect to $m$-reflection and anti-symmetric with respect to species interchange may be accessed from the ground state via an out-of-phase modulation of $g_{AA}$ and $g_{BB}$. In particular, of the five lowest excitations considered here, in the limit $g = \delta g$ only $|\Psi_2^{(g = \delta g)}\rangle$ may be accessed from the ground state via an out-of-phase modulation of the interaction parameters. This persists up to the mode crossing at $g\approx 1.3$, after which $|\Psi_3^{(g\gtrsim 1.3)}\rangle$ is the only state out of the five lowest excitations that is accessible from the ground state via an out-of-phase modulation of $g_{AA}$ and $g_{BB}$, in agreement with the computed transition matrix elements shown in Fig.~\ref{fig: Matrix Elements}(d). Thus $|\Psi_3^{(g)}\rangle$ is the lowest state accessible from the ground state through out-of-phase modulations of the interactions for $g\gtrsim 1.3$. 

\begin{figure}[h!]
    \centering
    \includegraphics[width=8cm,trim={2.5cm 7cm 5.5cm 8cm},clip]{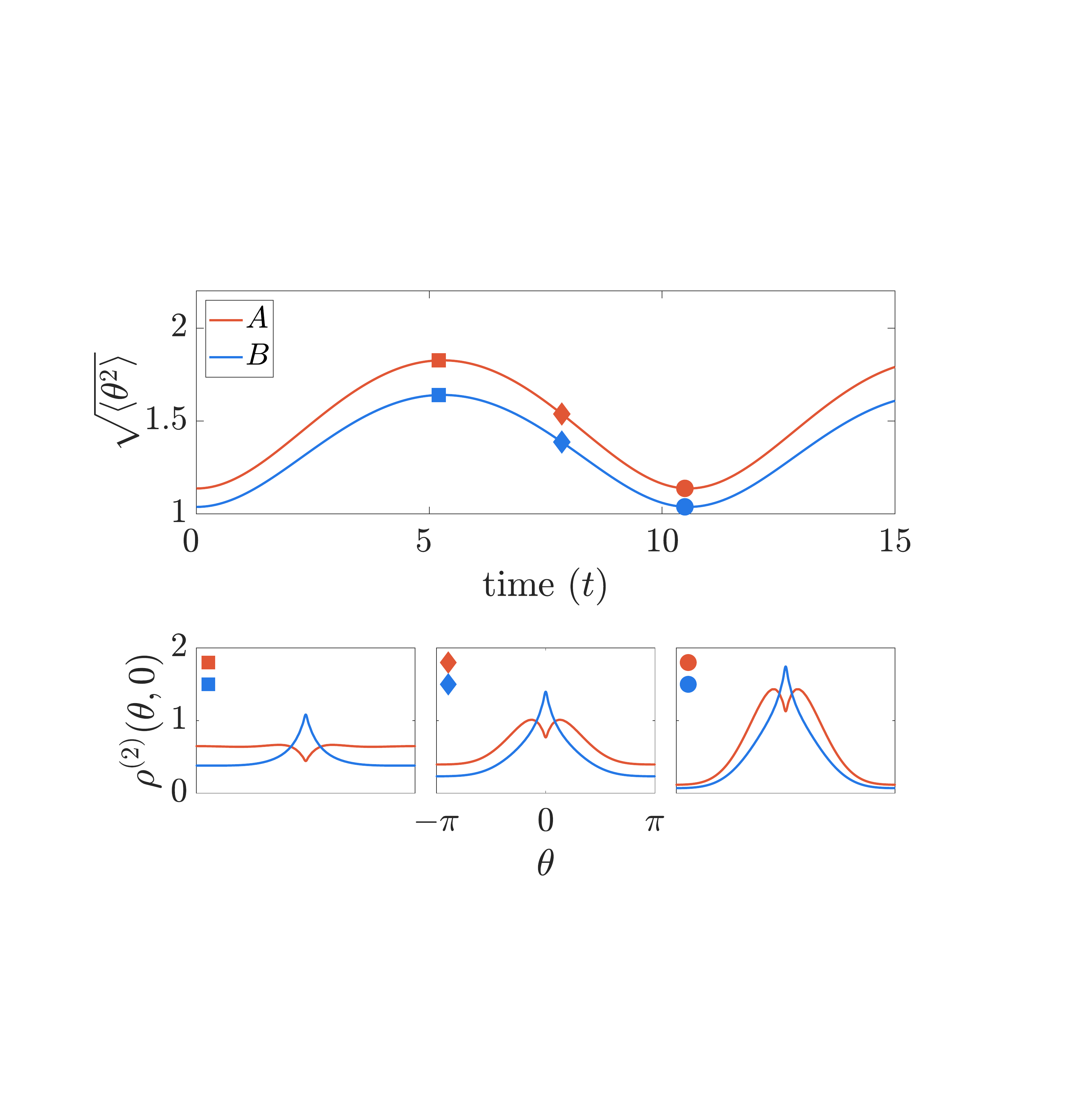}
    \caption{\textit{Upper panel:} {Mean angular width} $\sqrt{\langle \theta ^2\rangle }$ as a function of time for the superposition of the ground state  and first excited state, $\frac{1}{\sqrt{2}}(e^{-iE_0 t}|\Psi_0^{(g)}\rangle +e^{-E_1 t}|\Psi_1^{(g)}\rangle)$ at $g = 2.0$ and $g_{AB} = -0.9g$ in the $L=0$ subspace. 
    \textit{Lower panel:} 
     Pair correlations $\rho^{(2)}_{AA}(\theta,0)$  (red) and $\rho^{(2)}_{BA}(\theta,0)$ (blue) corresponding to the first (left) second (center) and third (right) pairs of time-points indicated by the corresponding markers in both panels.}    
    \label{fig:BreathingA}
\end{figure}

Periodic modulation of the interaction strengths may also produce a superposition of the many-body ground state and an accessible excited state 
\begin{equation}
    |\Psi (t)\rangle = \frac{1}{\sqrt{2}}e^{-iE_0 t}|\Psi^{(g)}_0\rangle + \frac{1}{\sqrt{2}}e^{-iE_i t}|\Psi^{(g)}_i\rangle
\end{equation}
(here neglecting higher-order couplings for simplicity). 
In Fig.~\ref{fig:BreathingA} we present such a superposition of the ground state $|\Psi_0^{(g)}\rangle$ and the lowest HA-like mode 
$|\Psi_1^{(g)}\rangle$ 
at $g=2$ and $g_{AB} = -0.9g$, 
which may be prepared from the ground state via an in-phase periodic modulation of $g_{AA}$ and $g_{BB}$ or via a periodic modulation of $g_{AB}$ such as in Eq.~\eqref{eqn: modulation of gAB}.
The upper panel shows the time evolution of the mean angular widths, 
\begin{equation}
\sqrt {\langle \theta ^2 \rangle} = \biggl (\int _{-\pi }^{\pi }\rho ^{(2)}_{\sigma  \sigma '}(\theta, 0) \theta ^2 {\mathrm{d}}\theta \biggl / \biggr. \int  _{-\pi }^{\pi }\rho ^{(2)}_{\sigma \sigma '}(\theta, 0){\mathrm{d}}\theta \biggr  )^{1/2}~,
\nonumber
\end{equation}
and the lower panel shows the pair correlations for each species at various time-points.
\begin{figure}[h!]
    \centering
    \includegraphics[width=8cm,trim={2.5cm 7cm 5.5cm 8cm},clip]{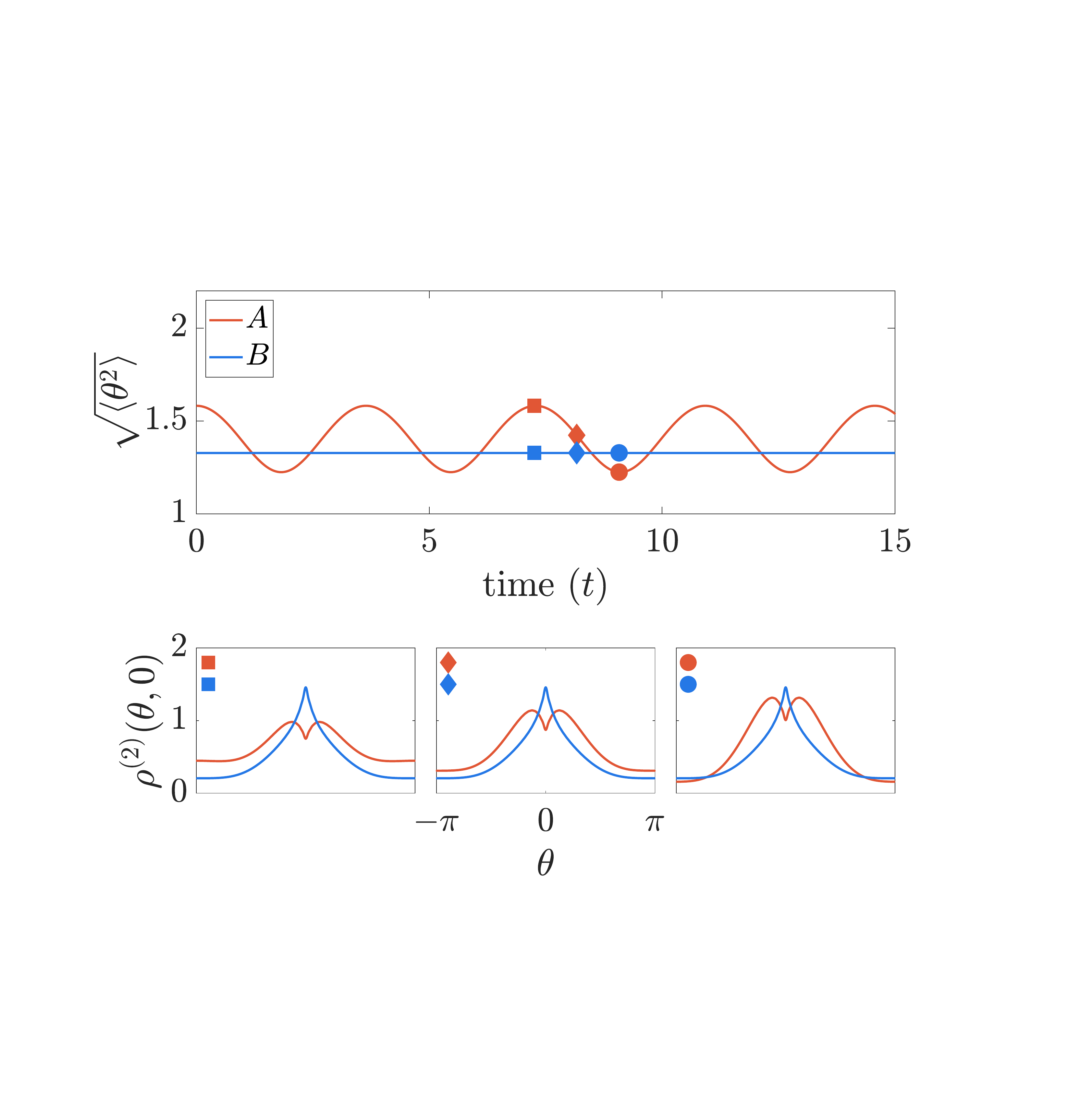}
    \caption{As in Fig.~\ref{fig:BreathingA} but now  for the superposition of the ground state and third excited state,  
     $\frac{1}{\sqrt{2}}(e^{-iE_0 t}|\Psi_0^{(g)}\rangle + e^{-iE_3 t}|\Psi_3^{(g)}\rangle)$,  in the $L=0$ subspace.}
    \label{fig:BreathingB}
\end{figure}
The two components expand and contract in phase with one another. 
Thus the nature of the breathing mode dynamics  reflects the fact that the considered superposition can be obtained by varying $g_{AA}$ and $g_{BB}$ in phase with one another. This in itself is a reflection of the fact that both
$|\Psi_1^{(g=2)}\rangle$ and $|\Psi_0^{(g=2)}\rangle$
are symmetric with respect to both system symmetries.
In Fig.~\ref{fig:BreathingB} we consider a superposition of the ground state $|\Psi_0^{(g)}\rangle$ and $|\Psi_3^{(g)}\rangle$ at $g=2$ and $g_{AB}=-0.9g$.
Now we observe a breathing mode primarily in one component (see the upper panel of Fig.~\ref{fig:BreathingB}), reflecting the fact that such a superposition is created by the out-of-phase modulation of $g_{AA}$ and $g_{BB}$ and that $|\Psi_3^{(g=2)}\rangle$ 
is anti-symmetric with respect to species interchange.

\section{Summary and Outlook}
\label{sec:Summary}

In summary, an ultra-cold binary bosonic mixture on a one dimensional ring has a homogeneous and a localized droplet phase.
Here, we have studied the few-body properties of these mixtures in the homogeneous-to-droplet crossover region. 
By varying the intra-species repulsion and inter-species attraction we found signatures of the few-body phase transition in the zero angular momentum ground state pair correlations, in the rotational properties of the low lying exact energy spectra and in the low-lying excitation modes. Increasing the inter-species attraction for fixed intra-species repulsion caused the pair correlations to change from a homogeneous to a localized distribution on the ring. 
It likewise drove a change in the yrast line, from the negative curvature associated with persistent currents, to a parabolic curvature indicative of rigid body rotation. 
The simultaneous onset of both phenomena clearly suggests the formation of a localized state. The consistency of these exact results with the phase transition predicted by the eGP approach is noteworthy. In the exact low-lying excitation spectra for fixed $g_{AB}/g$ the formation of a localized state manifested as a set of rotational modes and non-monotonic behavior in the lowest zero angular momentum mode around the point of phase transition.
We further analyzed the zero angular momentum excitations in terms of their behavior under the transformations which define two global symmetries of the system. This gave insight into the transition matrix elements and breathing mode dynamics of the excitations when in superposition with the many-body ground state.
We found in-phase breathing modes that are captured by most BMF treatments of the droplet problem as well as out-of-phase oscillations. We saw that the symmetry properties of the lowest lying modes in the limit of perturbatively weak interactions reflects the breathing mode dynamics of their associated states in the droplet phase in superposition with the ground state.
In the excitations for fixed $g$ and variable $g_{AB}$ we similarly see the formation of rotational modes  and a minimum in the lowest zero angular momentum mode. This hints  towards a spontaneously broken translation symmetry associated with few-body precursors to the collective Higgs-Andersson-like amplitude  and Goldstone-like phase  modes.  A similar phenomenology was found for systems with $N=10$ and 12 bosons, albeit with a reduced energetic convergence due to the significantly increased numerical effort at larger particle numbers. 

In  outlook to future work, new perspectives  will arise from studies of mass- and atom-number imbalanced mixtures, approaching a limit where one of the components may act as an embedded impurity supporting a many-boson bound state~\cite{Brauneis2022}. It will be intriguing to see how collective modes are modified when transitioning from the balanced symmetric case to only a single atom in one component. 
It was  recently shown that a  mobile impurity in a hetero-nuclear bosonic mixture may induce the system to localize into a droplet phase, 
with important prospects for the spectroscopic investigation of quantum fluctuations in a few-body environment~\cite{Bighin2022}.
Depending on the strength of inter- and intra-component interactions, an interesting question is how an imbalance or the presence of an impurity will affect the onset of symmetry breaking and the elementary modes signaling it. We expect experiments to be capable of reaching the few-body limit with bosons and fermions alike, opening up new avenues for a better bottom-up understanding of phase transitions, atom by atom. 

\begin{acknowledgments}
We thank M. Nilsson Tengstrand for discussions and valuable input on the solutions in the extended Gross-Pitaevskii approach. We also thank K. Mukherjee and T. Arnone Cardinale for discussions. 
This research was financially supported by the Knut and Alice Wallenberg Foundation, the Swedish Research Council and NanoLund.

\end{acknowledgments}

\appendix
\section{Hilbert Space Construction and Convergence Data}
\label{sec: Appendix ITCI}
We have employed a so called importance-truncated configuration interaction (ITCI) method~\cite{Roth2009}. The angular momentum eigenfunctions
$\phi_m(\theta) = \frac{1}{\sqrt{2\pi}}e^{im\theta}$
are used for the one-body basis, where integer $|m| \leq m_\text{max} = 60$ is the one-body angular momentum quantum number and $\theta$ is the azimuthal position on the ring. 
In the ITCI method the full many-body Hilbert space $\mathcal{H}$ is divided into a reference subspace $\mathcal{H}_\text{ref}$ in which diagonalization is performed and an orthogonal complementary subspace $\mathcal{H}_C$. We begin with a {small} reference subspace (spanned by many-body basis states with energy less than $10$)
in which the target energy eigenstate $|\psi_\text{ref}\rangle$ is {constructed}. 
\begin{figure}[h!]
    \centering
    \includegraphics[width=8cm,trim={0cm 0cm 0cm 0cm},clip]{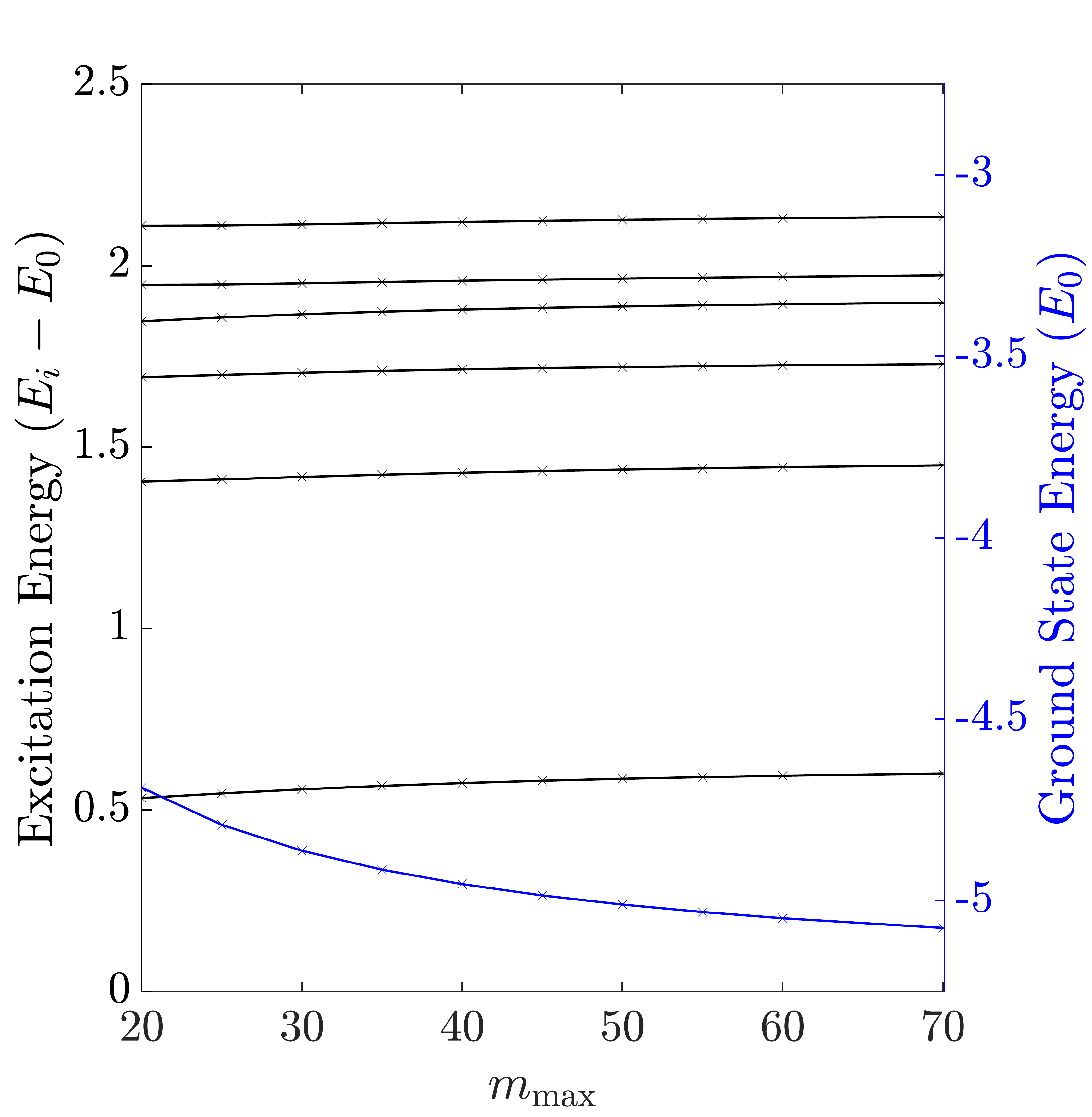}
    \caption{Convergence of the six lowest excitation energies ${E_i - E_0}$ (black) and the ground state energy $E_0$ (blue) as a function of $m_\text{max}$, where the single particle basis size is $2m_\text{max}+1 $, for a system of ${N=8}$ particles with $g=2$, $g_{AB} = -1.8$ and total angular momentum $L=0$. For weaker interaction strengths more rapid convergence is typically seen.
}
    \label{fig:Conv}
\end{figure}

$\mathcal{H}_\text{ref}$ is iteratively updated by transferring relevant states from $\mathcal{H}_C$ to $\mathcal{H}_\text{ref}$ to improve the target state until a desired accuracy is reached. In each iteration the target state is first constructed in $\mathcal{H}_\text{ref}$ and then expanded in $\mathcal{H}_C$ via multi-configurational first order perturbation theory with Epstein-Nesbet-like partitioning {\cite{Epstein1926,Nesbet1955}. For a detailed description of these methods see {\it e.g.}, Ref. \cite{Roth2009} and for selected configuration interaction methods more generally see {\it e.g.}, Ref. \cite{Tubman2020}. 
In this work the importance measure $\kappa_\nu$ for each Fock state $|\phi_\nu\rangle \in \mathcal{H}_C$ is taken to be its dimensionless perturbative amplitude in the expansion of $|\psi_\text{ref}\rangle$ and the importance threshold is $\kappa_\text{min} = 10^{-5}$. In each iteration all states with $\kappa_\nu > \kappa_\text{min}$ are transferred from $\mathcal{H}_C$ to $\mathcal{H}_\text{ref}$. In this way $\mathcal{H}_\text{ref}$ is tailored to the target state and the system Hamiltonian. To reduce computational time, reference threshold $C_\text{min} = 10^{-4}$ is used. That is, only states in $\mathcal{H}_\text{ref}$ with amplitude greater than $C_\text{min} = 10^{-4}$ are included in the reference state $|\psi_\text{ref}\rangle$ used to calculate the importance measures. The iterative search for relevant basis states is terminated when the relative energy difference of the target state between two iterations is less than $10^{-5}$.

The overall convergence obtained in this scheme is depicted in Fig.~\ref{fig:Conv}, which shows the ground state energy $E_0$ and the six lowest excitation energies ${E_i - E_0}$, $1\le i\le 6$, as a function of increasing single-particle basis size with one-body angular momentum cutoff $|m| \leq m_\text{max} $.

For completeness and future comparison with other approaches, the Table to the right lists the 
ground state energies and excitation energies obtained for the single-particle basis with $|m| \leq m_\text{max} = 60$
(see the data plotted in Fig.~\ref{fig:SpectraA}).

\begin{table}
\bigskip
\begin{center}
\resizebox{0.48\textwidth}{!}{%
\begin{tabular}{  c | c | c | c | c | c | c  }
    \hline\hline
$g$ & $E_0$  &$E_1 - E_0$ &$E_2 - E_0$ &$E_3 - E_0$ &$E_4 - E_0$ &$E_5 - E_0$\\
    \hline
    ~~~0.0100~~~~&   ~~~-0.0040~~~&    ~~~0.9981~~~&    ~~~1.0079~~~&    ~~~1.0109~~~&    ~~~1.0179~~~&    ~~~1.9963~~~\\
    0.0200&   -0.0084&    0.9963&    1.0158&    1.0216&    1.0357&    1.9927\\
    0.0500&   -0.0240&    0.9907&    1.0389&    1.0528&    1.0884&    1.9820\\
    0.1000&   -0.0567&    0.9813&    1.0759&    1.1014&    1.1732&    1.9645\\
    0.2000&   -0.1444&    0.9616&    1.1431&    1.1876&    1.3309&    1.9291\\
    0.3000&   -0.2573&    0.9400&    1.2018&    1.2615&    1.4730&    1.8925\\
    0.4000&   -0.3920&    0.9159&    1.2527&    1.3254&    1.6007&    1.8542\\
    0.5000&   -0.5464&    0.8893&    1.2967&    1.3809&    1.7143&    1.8154\\
    0.6000&   -0.7189&    0.8602&    1.3345&    1.4290&    1.7647&    1.8261\\
    0.7000&   -0.9086&    0.8285&    1.3668&    1.4708&    1.7244&    1.9142\\
    0.8000&   -1.1150&    0.7945&    1.3942&    1.5071&    1.6799&    1.9927\\
    0.9000&   -1.3376&    0.7582&    1.4174&    1.5384&    1.6345&    2.0568\\
    1.0000&   -1.5765&    0.7202&    1.4370&    1.5655&    1.5892&    2.1024\\
    1.1000&   -1.8318&    0.6809&    1.4540&    1.5447&    1.5891&    2.1246\\
    1.2000&   -2.1039&    0.6412&    1.4691&    1.5021&    1.6102&    2.1209\\
    1.3000&   -2.3937&    0.6027&    1.4628&    1.4836&    1.6300&    2.0974\\
    1.4000&   -2.7022&    0.5671&    1.4285&    1.4990&    1.6499&    2.0646\\
    1.5000&   -3.0312&    0.5375&    1.4009&    1.5171&    1.6721&    2.0301\\
    1.6000&   -3.3824&    0.5176&    1.3829&    1.5405&    1.6990&    1.9996\\
    1.7000&   -3.7583&    0.5109&    1.3765&    1.5714&    1.7328&    1.9761\\
    1.8000&   -4.1607&    0.5207&    1.3840&    1.6118&    1.7758&    1.9624\\
    1.9000&   -4.5907&    0.5487&    1.4066&    1.6630&    1.8293&    1.9600\\
    2.0000&   -5.0484&    0.5944&    1.4445&    1.7252&    1.8936&    1.9693\\
    2.1000&   -5.5331&    0.6558&    1.4971&    1.7980&    1.9683&    1.9897\\
    2.2000&   -6.0439&    0.7302&    1.5633&    1.8803&    2.0206&    2.0525\\
    2.3000&   -6.5795&    0.8148&    1.6419&    1.9712&    2.0609&    2.1452\\
    2.4000&   -7.1390&    0.9072&    1.7314&    2.0698&    2.1098&    2.2455\\
    2.5000&   -7.7216&    1.0054&    1.8307&    2.1669&    2.1753&    2.3282\\
    \hline\hline
\end{tabular}
}
\end{center}
\caption{Table of ground state energies $E_0$ and excitation energies $E_i-E_0$ for 
$1\le i\le 5$ with $g_{AB}=-0.9g$, $N=8$ and $L=0$, for a single-particle basis of $|m| \leq m_\text{max} = 60$.
}
\end{table}
\bigskip

\section{Bogoliubov results and stability of the homogeneous solution}
\label{sec:Appendix}

Following the common Bogoliubov procedure, the Hamiltonian Eq.(\ref{eqn:ExactHamiltonian}) takes the usual quadratic form. We introduce the new operators $\hat{b}_{\sigma,m}$, defined by the transformation
\begin{equation}\label{eqn: bogoliubov transformation}
    \begin{bmatrix}
        \hat{a}_{A,m} \\
        \hat{a}_{A,-m}^{\dagger} \\
        \hat{a}_{B,m}\\
        \hat{a}_{B,-m}^{\dagger}
    \end{bmatrix}
    =
    \begin{bmatrix}
        u_{1,m} & v_{1,m} & \mu_{1,m} & \nu_{1,m} \\
        v_{1,m} & u_{1,m} & \nu_{1,m} & \mu_{1,m} \\
        u_{2,m} & v_{2,m} & \mu_{2,m} & \nu_{2,m} \\
        v_{2,m} & u_{2,m} & \nu_{2,m} & \mu_{2,m} \\
    \end{bmatrix}
    \begin{bmatrix}
        \hat{b}_{A,m} \\
        \hat{b}_{A,-m}^{\dagger} \\
        \hat{b}_{B,m}\\
        \hat{b}_{B,-m}^{\dagger}
    \end{bmatrix},
\end{equation}
and impose the standard constraints of the commutation relations $[\hat{a}_{\sigma,\pm m}, \hat{a}_{\sigma,\pm m}^{\dagger}] = 1$ and $[\hat{a}_{A,m}, \hat{a}_{B,m}] = [\hat{a}_{A,m}, \hat{a}_{B,m}^{\dagger}] = 0$. Four additional constraints are obtained by imposing the condition that the Hamiltonian is diagonal in the new operators. 
We then have eight constraints and eight unknowns for each {mode} $m$, and are thus able to evaluate the matrix elements which appear in the transformation of Eq.~\eqref{eqn: bogoliubov transformation}. The result is the diagonal Hamiltonian
($\hbar = M = R = 1$) 
\begin{equation}\label{eqn: diagonal Hamiltonian}
    \begin{split}
        \hat{H}  = & E_0 + E_\text{LHY}\\
        &+ \sum_{m > 0} \Big[ \sqrt{e_m [n_0 (g - g_{AB}) + e_m]}\\
        & \quad\quad\quad\quad\quad\times(\hat{b}_{A,m}^{\dagger} \hat{b}_{A,m} + \hat{b}_{A,-m}^{\dagger} \hat{b}_{A,-m}) \\
        & + \sqrt{(e_m [n_0 (g + g_{AB}) + e_m]}\\
        &\quad\quad\quad\quad\quad\times(b_{B,m}^{\dagger} \hat{b}_{B,m} + \hat{b}_{B,-m}^{\dagger} \hat{b}_{B,-m})\Big]\\
    \end{split}
\end{equation}
where $n_0 = N/2\pi $, $e_m=m^2/2$ and we have imposed the simplifying conditions of equal particle numbers and equal intra-species interactions $g = g_{AA} = g_{BB}$. Furthermore, 
\begin{equation}\label{eqn: E_0}
        E_0= \frac{1}{4}(g + g_{AB}) n_0 N 
\end{equation}
is the leading order, mean-field contribution to the energy and 
\begin{equation}\label{eqn: total energy density}
    \begin{split}         
        E_\text{LHY} = \sum_{m \neq 0} \frac{1}{2}\Big[ &\sqrt{e_m[n_0 (g - g_{AB}) + e_m]} \\
        & + \sqrt{e_m [n_0 (g + g_{AB}) + e_m]}  \Big] \\
        & - n_0 g/2 - e_m
    \end{split}
\end{equation}
is the next-order LHY correction. 
Converting this sum to an integral and dividing by the length of the system, $l$, we find the energy per unit length  of the ground state
\begin{equation}\label{eqn: total energy denisty}
    (E_0 + E_\text{LHY})/l = \frac{1}{2}g_1 n_0^2 - \frac{2}{3}g_2n_0^{3/2}
\end{equation}
where we have introduced the parameters
\begin{equation}\label{eqn: g1 and g2}
    \begin{split}
        &g_1 = \frac{1}{2}(g + g_{AB})\\
        &g_2 = \frac{1}{2^{3/2} \pi}[(g + g_{AB})^{3/2} + (g - g_{AB})^{3/2}].
    \end{split}
\end{equation}
Clearly, for $g \approx -g_{AB}$ the mean-field energy is small and the next-order LHY contribution becomes significant. In particular, since $E_\text{LHY}$ is always negative in one dimension, for sufficiently small and positive $g_1$,
$E_0$ and $E_\text{LHY}$ may balance. As the two terms scale differently in the density, one may then expect to find localized bound states stabilized by the attractive quantum fluctuations~\cite{Petrov2016}.
This observation motivates our choice of interaction parameters 
for the exact calculations presented above. In particular we consider attractive inter-species interactions $g_{AB} < 0$ and repulsive intra-species interactions $g > |g_{AB}|$.

For equal particle numbers and equal intra-species interactions the beyond mean-field Hamiltonian can be expressed in terms of a single order parameter $\Psi(\theta)$. In particular, from Eq.~\eqref{eqn: total energy denisty} it follows that
\begin{equation}\label{eqn: Mean-field Hamiltonian}
    \begin{split}
        \mathcal{E} = \frac{N}{2}\int &\left|\Psi_\theta \right|^2 d\theta + \frac{1}{2}g_1 N^2\int \left|\Psi\right|^4 d\theta \\
        &-\frac{2}{3}g_2 N^{3/2}\int \left|\Psi \right|^3 d\theta.
    \end{split}
\end{equation}
where $\Psi(\theta)$ satisfies Eq.~\eqref{eqn: order parameter diff eqn}, which we repeat here for convenience, 
\begin{equation}\label{eqn: order parameter diff eqn 2}
    \begin{split}
        \mu\Psi = &-\frac{1}{2}\frac{\partial^2\Psi}{\partial \theta^2} + {g_1 N}\left|\Psi\right|^2 \Psi
        - {g_2 \sqrt{N}}\left|\Psi\right|\Psi
    \end{split}
\end{equation}
in terms of $g_1$ and $g_2$ as defined in Eq.~\eqref{eqn: g1 and g2}, and $\int|\Psi(\theta)|^2 d\theta = 1$. 

A trivial solution to Eq.~\eqref{eqn: order parameter diff eqn 2} is the homogeneous solution $\Psi_0$, however this solution is not always stable.
%
%%%%%%%%%%%%%%%%%%%%%%%%%%%%%%%%% Dynamic stability %%%%%%%%%%%%%%%%%%%%%%%%%%%%%%%%%%%%%%%%%%%%%%%%%%%%%%%%%%%%%%%%%%%%%%%%%%%
To examine the dynamic stability of this state with homogeneous density and periodic boundary conditions,
consider an order parameter that has only small deviations from the homogeneous state $\Tilde{\Psi} = \Psi_0 + \delta\Psi$. The deviations $\delta \Psi$ are governed by the following equation
\begin{equation}\label{eqn: delta psi diff eqn }
    \begin{split}
        i\frac{\partial \delta\psi}{\partial t} = - &\frac{1}{2} \frac{\partial^2 \delta \Psi}{\partial \theta^2} \\
        &+ \Big(g_1 n_0 - \frac{1}{2}g_2 \sqrt{n_0}\Big) [\delta\Psi + (\delta\Psi)^*].
    \end{split}
\end{equation}
By assuming plane-wave solutions, $\delta\Psi \propto e^{im\theta - i\omega t}$, we find that the dispersion relation takes the form
\begin{equation}
    \begin{split}
        \omega &= \sqrt{e_m \Big(2g_1 n_0 - g_2\sqrt{n_0} + e_m\Big)}
    \end{split}
\end{equation}
This equation gives a speed of sound that coincides with the one predicted by Bogoliubov, however it also includes the next-order correction. Furthermore, it implies an instability, with the most unstable mode corresponding to $m = 1$. Specifically, this dynamic instability occurs when
\begin{equation}\label{eqn: appendix stability}
    \begin{split}
        &0 > - \frac{1}{2}g_2 \sqrt{n_0}+ g_1 n_0 + \frac{1}{4}.\\
    \end{split}
\end{equation}
Eq.~\eqref{eqn: appendix stability} defines the phase boundary between the homogeneous and localized phases (see also Eq.~\ref{eqn:instability} in the main text).

The condition for the energetic stability of the homogeneous solution yields an identical result.
This can be shown by considering the order parameter
\begin{equation}
    \Psi = \frac{1}{\sqrt{2\pi}}(c_0 + 2c_1\cos\theta )
\end{equation}
where $c_0^2 + 2c_1^2 = 1$. Assuming that $|c_0| \gg |c_1|$ and expanding the energy of the system up to second order in $c_1$, we see that the kinetic energy is trivially $c_1^2$, the energy from the contact term is $4g_1 n_0 c_1^2$ and finally the BMF term is $-2g_2\sqrt{n_0}c_1^2$. Therefore, in order for the homogeneous solution to be energetically unstable, we end up with the same condition that was derived from the condition for dynamic stability, Eq.~\eqref{eqn: appendix stability}.

\bigskip

%LIBRARY
%\bibliography{Chergui_Project1.bib}

\begin{thebibliography}{88}%
\makeatletter
\providecommand \@ifxundefined [1]{%
 \@ifx{#1\undefined}
}%
\providecommand \@ifnum [1]{%
 \ifnum #1\expandafter \@firstoftwo
 \else \expandafter \@secondoftwo
 \fi
}%
\providecommand \@ifx [1]{%
 \ifx #1\expandafter \@firstoftwo
 \else \expandafter \@secondoftwo
 \fi
}%
\providecommand \natexlab [1]{#1}%
\providecommand \enquote  [1]{``#1''}%
\providecommand \bibnamefont  [1]{#1}%
\providecommand \bibfnamefont [1]{#1}%
\providecommand \citenamefont [1]{#1}%
\providecommand \href@noop [0]{\@secondoftwo}%
\providecommand \href [0]{\begingroup \@sanitize@url \@href}%
\providecommand \@href[1]{\@@startlink{#1}\@@href}%
\providecommand \@@href[1]{\endgroup#1\@@endlink}%
\providecommand \@sanitize@url [0]{\catcode `\\12\catcode `\$12\catcode
  `\&12\catcode `\#12\catcode `\^12\catcode `\_12\catcode `\%12\relax}%
\providecommand \@@startlink[1]{}%
\providecommand \@@endlink[0]{}%
\providecommand \url  [0]{\begingroup\@sanitize@url \@url }%
\providecommand \@url [1]{\endgroup\@href {#1}{\urlprefix }}%
\providecommand \urlprefix  [0]{URL }%
\providecommand \Eprint [0]{\href }%
\providecommand \doibase [0]{http://dx.doi.org/}%
\providecommand \selectlanguage [0]{\@gobble}%
\providecommand \bibinfo  [0]{\@secondoftwo}%
\providecommand \bibfield  [0]{\@secondoftwo}%
\providecommand \translation [1]{[#1]}%
\providecommand \BibitemOpen [0]{}%
\providecommand \bibitemStop [0]{}%
\providecommand \bibitemNoStop [0]{.\EOS\space}%
\providecommand \EOS [0]{\spacefactor3000\relax}%
\providecommand \BibitemShut  [1]{\csname bibitem#1\endcsname}%
\let\auto@bib@innerbib\@empty
%</preamble>
\bibitem [{\citenamefont {Bulgac}(2002)}]{Bulgac2002}%
  \BibitemOpen
  \bibfield  {author} {\bibinfo {author} {\bibfnamefont {A.}~\bibnamefont
  {Bulgac}},\ }\href {\doibase 10.1103/PhysRevLett.89.050402} {\bibfield
  {journal} {\bibinfo  {journal} {Phys. Rev. Lett.}\ }\textbf {\bibinfo
  {volume} {89}},\ \bibinfo {pages} {050402} (\bibinfo {year}
  {2002})}\BibitemShut {NoStop}%
\bibitem [{\citenamefont {Bedaque}\ \emph {et~al.}(2003)\citenamefont
  {Bedaque}, \citenamefont {Bulgac},\ and\ \citenamefont
  {Rupak}}]{Bedaque2003}%
  \BibitemOpen
  \bibfield  {author} {\bibinfo {author} {\bibfnamefont {P.~F.}\ \bibnamefont
  {Bedaque}}, \bibinfo {author} {\bibfnamefont {A.}~\bibnamefont {Bulgac}}, \
  and\ \bibinfo {author} {\bibfnamefont {G.}~\bibnamefont {Rupak}},\ }\href
  {\doibase 10.1103/PhysRevA.68.033606} {\bibfield  {journal} {\bibinfo
  {journal} {Phys. Rev. A}\ }\textbf {\bibinfo {volume} {68}},\ \bibinfo
  {pages} {033606} (\bibinfo {year} {2003})}\BibitemShut {NoStop}%
\bibitem [{\citenamefont {Hammer}\ and\ \citenamefont
  {Son}(2004)}]{Hammer2004}%
  \BibitemOpen
  \bibfield  {author} {\bibinfo {author} {\bibfnamefont {H.-W.}\ \bibnamefont
  {Hammer}}\ and\ \bibinfo {author} {\bibfnamefont {D.~T.}\ \bibnamefont
  {Son}},\ }\href {\doibase 10.1103/PhysRevLett.93.250408} {\bibfield
  {journal} {\bibinfo  {journal} {Phys. Rev. Lett.}\ }\textbf {\bibinfo
  {volume} {93}},\ \bibinfo {pages} {250408} (\bibinfo {year}
  {2004})}\BibitemShut {NoStop}%
\bibitem [{\citenamefont {Lee}\ \emph {et~al.}(1957)\citenamefont {Lee},
  \citenamefont {Huang},\ and\ \citenamefont {Yang}}]{Lee1957}%
  \BibitemOpen
  \bibfield  {author} {\bibinfo {author} {\bibfnamefont {T.~D.}\ \bibnamefont
  {Lee}}, \bibinfo {author} {\bibfnamefont {K.}~\bibnamefont {Huang}}, \ and\
  \bibinfo {author} {\bibfnamefont {C.~N.}\ \bibnamefont {Yang}},\ }\href
  {\doibase 10.1103/PhysRev.106.1135} {\bibfield  {journal} {\bibinfo
  {journal} {Phys. Rev.}\ }\textbf {\bibinfo {volume} {106}},\ \bibinfo {pages}
  {1135} (\bibinfo {year} {1957})}\BibitemShut {NoStop}%
\bibitem [{\citenamefont {Petrov}(2015)}]{Petrov2015}%
  \BibitemOpen
  \bibfield  {author} {\bibinfo {author} {\bibfnamefont {D.~S.}\ \bibnamefont
  {Petrov}},\ }\href {\doibase 10.1103/PhysRevLett.115.155302} {\bibfield
  {journal} {\bibinfo  {journal} {Phys. Rev. Lett.}\ }\textbf {\bibinfo
  {volume} {115}},\ \bibinfo {pages} {155302} (\bibinfo {year}
  {2015})}\BibitemShut {NoStop}%
\bibitem [{\citenamefont {Petrov}\ and\ \citenamefont
  {Astrakharchik}(2016)}]{Petrov2016}%
  \BibitemOpen
  \bibfield  {author} {\bibinfo {author} {\bibfnamefont {D.~S.}\ \bibnamefont
  {Petrov}}\ and\ \bibinfo {author} {\bibfnamefont {G.~E.}\ \bibnamefont
  {Astrakharchik}},\ }\href {\doibase 10.1103/PhysRevLett.117.100401}
  {\bibfield  {journal} {\bibinfo  {journal} {Phys. Rev. Lett.}\ }\textbf
  {\bibinfo {volume} {117}},\ \bibinfo {pages} {100401} (\bibinfo {year}
  {2016})}\BibitemShut {NoStop}%
\bibitem [{\citenamefont {Kadau}\ \emph {et~al.}(2016)\citenamefont {Kadau},
  \citenamefont {Schmitt}, \citenamefont {Wenzel}, \citenamefont {Wink},
  \citenamefont {Maier}, \citenamefont {Ferrier-Barbut},\ and\ \citenamefont
  {Pfau}}]{Kadau2016}%
  \BibitemOpen
  \bibfield  {author} {\bibinfo {author} {\bibfnamefont {H.}~\bibnamefont
  {Kadau}}, \bibinfo {author} {\bibfnamefont {M.}~\bibnamefont {Schmitt}},
  \bibinfo {author} {\bibfnamefont {M.}~\bibnamefont {Wenzel}}, \bibinfo
  {author} {\bibfnamefont {C.}~\bibnamefont {Wink}}, \bibinfo {author}
  {\bibfnamefont {T.}~\bibnamefont {Maier}}, \bibinfo {author} {\bibfnamefont
  {I.}~\bibnamefont {Ferrier-Barbut}}, \ and\ \bibinfo {author} {\bibfnamefont
  {T.}~\bibnamefont {Pfau}},\ }\href {\doibase 10.1038/nature16485} {\bibfield
  {journal} {\bibinfo  {journal} {Nature}\ }\textbf {\bibinfo {volume} {530}},\
  \bibinfo {pages} {194} (\bibinfo {year} {2016})}\BibitemShut {NoStop}%
\bibitem [{\citenamefont {Ferrier-Barbut}\ \emph
  {et~al.}(2016{\natexlab{a}})\citenamefont {Ferrier-Barbut}, \citenamefont
  {Kadau}, \citenamefont {Schmitt}, \citenamefont {Wenzel},\ and\ \citenamefont
  {Pfau}}]{Ferrier-Barbut2016}%
  \BibitemOpen
  \bibfield  {author} {\bibinfo {author} {\bibfnamefont {I.}~\bibnamefont
  {Ferrier-Barbut}}, \bibinfo {author} {\bibfnamefont {H.}~\bibnamefont
  {Kadau}}, \bibinfo {author} {\bibfnamefont {M.}~\bibnamefont {Schmitt}},
  \bibinfo {author} {\bibfnamefont {M.}~\bibnamefont {Wenzel}}, \ and\ \bibinfo
  {author} {\bibfnamefont {T.}~\bibnamefont {Pfau}},\ }\href {\doibase
  10.1103/PhysRevLett.116.215301} {\bibfield  {journal} {\bibinfo  {journal}
  {Phys. Rev. Lett.}\ }\textbf {\bibinfo {volume} {116}},\ \bibinfo {pages}
  {215301} (\bibinfo {year} {2016}{\natexlab{a}})}\BibitemShut {NoStop}%
\bibitem [{\citenamefont {Ferrier-Barbut}\ \emph
  {et~al.}(2016{\natexlab{b}})\citenamefont {Ferrier-Barbut}, \citenamefont
  {Schmitt}, \citenamefont {Wenzel}, \citenamefont {Kadau},\ and\ \citenamefont
  {Pfau}}]{Ferrier-Barbut2016b}%
  \BibitemOpen
  \bibfield  {author} {\bibinfo {author} {\bibfnamefont {I.}~\bibnamefont
  {Ferrier-Barbut}}, \bibinfo {author} {\bibfnamefont {M.}~\bibnamefont
  {Schmitt}}, \bibinfo {author} {\bibfnamefont {M.}~\bibnamefont {Wenzel}},
  \bibinfo {author} {\bibfnamefont {H.}~\bibnamefont {Kadau}}, \ and\ \bibinfo
  {author} {\bibfnamefont {T.}~\bibnamefont {Pfau}},\ }\href {\doibase
  10.1088/0953-4075/49/21/214004} {\bibfield  {journal} {\bibinfo  {journal}
  {Journal of Physics B: Atomic, Molecular and Optical Physics}\ }\textbf
  {\bibinfo {volume} {49}},\ \bibinfo {pages} {214004} (\bibinfo {year}
  {2016}{\natexlab{b}})}\BibitemShut {NoStop}%
\bibitem [{\citenamefont {Schmitt}\ \emph {et~al.}(2016)\citenamefont
  {Schmitt}, \citenamefont {Wenzel}, \citenamefont {B{\"o}ttcher},
  \citenamefont {Ferrier-Barbut},\ and\ \citenamefont {Pfau}}]{Schmitt2016}%
  \BibitemOpen
  \bibfield  {author} {\bibinfo {author} {\bibfnamefont {M.}~\bibnamefont
  {Schmitt}}, \bibinfo {author} {\bibfnamefont {M.}~\bibnamefont {Wenzel}},
  \bibinfo {author} {\bibfnamefont {F.}~\bibnamefont {B{\"o}ttcher}}, \bibinfo
  {author} {\bibfnamefont {I.}~\bibnamefont {Ferrier-Barbut}}, \ and\ \bibinfo
  {author} {\bibfnamefont {T.}~\bibnamefont {Pfau}},\ }\href {\doibase
  10.1038/nature20126} {\bibfield  {journal} {\bibinfo  {journal} {Nature}\
  }\textbf {\bibinfo {volume} {539}},\ \bibinfo {pages} {259} (\bibinfo {year}
  {2016})}\BibitemShut {NoStop}%
\bibitem [{\citenamefont {Chomaz}\ \emph {et~al.}(2016)\citenamefont {Chomaz},
  \citenamefont {Baier}, \citenamefont {Petter}, \citenamefont {Mark},
  \citenamefont {W\"achtler}, \citenamefont {Santos},\ and\ \citenamefont
  {Ferlaino}}]{Chomaz2016}%
  \BibitemOpen
  \bibfield  {author} {\bibinfo {author} {\bibfnamefont {L.}~\bibnamefont
  {Chomaz}}, \bibinfo {author} {\bibfnamefont {S.}~\bibnamefont {Baier}},
  \bibinfo {author} {\bibfnamefont {D.}~\bibnamefont {Petter}}, \bibinfo
  {author} {\bibfnamefont {M.~J.}\ \bibnamefont {Mark}}, \bibinfo {author}
  {\bibfnamefont {F.}~\bibnamefont {W\"achtler}}, \bibinfo {author}
  {\bibfnamefont {L.}~\bibnamefont {Santos}}, \ and\ \bibinfo {author}
  {\bibfnamefont {F.}~\bibnamefont {Ferlaino}},\ }\href {\doibase
  10.1103/PhysRevX.6.041039} {\bibfield  {journal} {\bibinfo  {journal} {Phys.
  Rev. X}\ }\textbf {\bibinfo {volume} {6}},\ \bibinfo {pages} {041039}
  (\bibinfo {year} {2016})}\BibitemShut {NoStop}%
\bibitem [{\citenamefont {W\"achtler}\ and\ \citenamefont
  {Santos}(2016{\natexlab{a}})}]{Wachtler2016a}%
  \BibitemOpen
  \bibfield  {author} {\bibinfo {author} {\bibfnamefont {F.}~\bibnamefont
  {W\"achtler}}\ and\ \bibinfo {author} {\bibfnamefont {L.}~\bibnamefont
  {Santos}},\ }\href {\doibase 10.1103/PhysRevA.94.043618} {\bibfield
  {journal} {\bibinfo  {journal} {Phys. Rev. A}\ }\textbf {\bibinfo {volume}
  {94}},\ \bibinfo {pages} {043618} (\bibinfo {year}
  {2016}{\natexlab{a}})}\BibitemShut {NoStop}%
\bibitem [{\citenamefont {W\"achtler}\ and\ \citenamefont
  {Santos}(2016{\natexlab{b}})}]{Wachtler2016b}%
  \BibitemOpen
  \bibfield  {author} {\bibinfo {author} {\bibfnamefont {F.}~\bibnamefont
  {W\"achtler}}\ and\ \bibinfo {author} {\bibfnamefont {L.}~\bibnamefont
  {Santos}},\ }\href {\doibase 10.1103/PhysRevA.93.061603} {\bibfield
  {journal} {\bibinfo  {journal} {Phys. Rev. A}\ }\textbf {\bibinfo {volume}
  {93}},\ \bibinfo {pages} {061603} (\bibinfo {year}
  {2016}{\natexlab{b}})}\BibitemShut {NoStop}%
\bibitem [{\citenamefont {Bisset}\ \emph {et~al.}(2016)\citenamefont {Bisset},
  \citenamefont {Wilson}, \citenamefont {Baillie},\ and\ \citenamefont
  {Blakie}}]{Bisset2016}%
  \BibitemOpen
  \bibfield  {author} {\bibinfo {author} {\bibfnamefont {R.~N.}\ \bibnamefont
  {Bisset}}, \bibinfo {author} {\bibfnamefont {R.~M.}\ \bibnamefont {Wilson}},
  \bibinfo {author} {\bibfnamefont {D.}~\bibnamefont {Baillie}}, \ and\
  \bibinfo {author} {\bibfnamefont {P.~B.}\ \bibnamefont {Blakie}},\ }\href
  {\doibase 10.1103/PhysRevA.94.033619} {\bibfield  {journal} {\bibinfo
  {journal} {Phys. Rev. A}\ }\textbf {\bibinfo {volume} {94}},\ \bibinfo
  {pages} {033619} (\bibinfo {year} {2016})}\BibitemShut {NoStop}%
\bibitem [{\citenamefont {Baillie}\ \emph {et~al.}(2017)\citenamefont
  {Baillie}, \citenamefont {Wilson},\ and\ \citenamefont
  {Blakie}}]{Baillie2017}%
  \BibitemOpen
  \bibfield  {author} {\bibinfo {author} {\bibfnamefont {D.}~\bibnamefont
  {Baillie}}, \bibinfo {author} {\bibfnamefont {R.~M.}\ \bibnamefont {Wilson}},
  \ and\ \bibinfo {author} {\bibfnamefont {P.~B.}\ \bibnamefont {Blakie}},\
  }\href {\doibase 10.1103/PhysRevLett.119.255302} {\bibfield  {journal}
  {\bibinfo  {journal} {Phys. Rev. Lett.}\ }\textbf {\bibinfo {volume} {119}},\
  \bibinfo {pages} {255302} (\bibinfo {year} {2017})}\BibitemShut {NoStop}%
\bibitem [{\citenamefont {Saito}(2016)}]{Saito2016}%
  \BibitemOpen
  \bibfield  {author} {\bibinfo {author} {\bibfnamefont {H.}~\bibnamefont
  {Saito}},\ }\href {\doibase 10.7566/JPSJ.85.053001} {\bibfield  {journal}
  {\bibinfo  {journal} {J. Phys. Soc. Jap.}\ }\textbf
  {\bibinfo {volume} {85}},\ \bibinfo {pages} {053001} (\bibinfo {year}
  {2016})}\BibitemShut {NoStop}%
\bibitem [{\citenamefont {Macia}\ \emph {et~al.}(2016)\citenamefont {Macia},
  \citenamefont {S\'anchez-Baena}, \citenamefont {Boronat},\ and\ \citenamefont
  {Mazzanti}}]{Macia2016}%
  \BibitemOpen
  \bibfield  {author} {\bibinfo {author} {\bibfnamefont {A.}~\bibnamefont
  {Macia}}, \bibinfo {author} {\bibfnamefont {J.}~\bibnamefont
  {S\'anchez-Baena}}, \bibinfo {author} {\bibfnamefont {J.}~\bibnamefont
  {Boronat}}, \ and\ \bibinfo {author} {\bibfnamefont {F.}~\bibnamefont
  {Mazzanti}},\ }\href {\doibase 10.1103/PhysRevLett.117.205301} {\bibfield
  {journal} {\bibinfo  {journal} {Phys. Rev. Lett.}\ }\textbf {\bibinfo
  {volume} {117}},\ \bibinfo {pages} {205301} (\bibinfo {year}
  {2016})}\BibitemShut {NoStop}%
\bibitem [{\citenamefont {Cinti}\ and\ \citenamefont
  {Boninsegni}(2017)}]{Cinti2017}%
  \BibitemOpen
  \bibfield  {author} {\bibinfo {author} {\bibfnamefont {F.}~\bibnamefont
  {Cinti}}\ and\ \bibinfo {author} {\bibfnamefont {M.}~\bibnamefont
  {Boninsegni}},\ }\href {\doibase 10.1103/PhysRevA.96.013627} {\bibfield
  {journal} {\bibinfo  {journal} {Phys. Rev. A}\ }\textbf {\bibinfo {volume}
  {96}},\ \bibinfo {pages} {013627} (\bibinfo {year} {2017})}\BibitemShut
  {NoStop}%
\bibitem [{\citenamefont {Cabrera}\ \emph {et~al.}(2018)\citenamefont
  {Cabrera}, \citenamefont {Tanzi}, \citenamefont {Sanz}, \citenamefont
  {Naylor}, \citenamefont {Thomas}, \citenamefont {Cheiney},\ and\
  \citenamefont {Tarruell}}]{Cabrera2018}%
  \BibitemOpen
  \bibfield  {author} {\bibinfo {author} {\bibfnamefont {C.~R.}\ \bibnamefont
  {Cabrera}}, \bibinfo {author} {\bibfnamefont {L.}~\bibnamefont {Tanzi}},
  \bibinfo {author} {\bibfnamefont {J.}~\bibnamefont {Sanz}}, \bibinfo {author}
  {\bibfnamefont {B.}~\bibnamefont {Naylor}}, \bibinfo {author} {\bibfnamefont
  {P.}~\bibnamefont {Thomas}}, \bibinfo {author} {\bibfnamefont
  {P.}~\bibnamefont {Cheiney}}, \ and\ \bibinfo {author} {\bibfnamefont
  {L.}~\bibnamefont {Tarruell}},\ }\href {\doibase 10.1126/science.aao5686}
  {\bibfield  {journal} {\bibinfo  {journal} {Science}\ }\textbf {\bibinfo
  {volume} {359}},\ \bibinfo {pages} {301} (\bibinfo {year}
  {2018})}\BibitemShut {NoStop}%
\bibitem [{\citenamefont {Semeghini}\ \emph {et~al.}(2018)\citenamefont
  {Semeghini}, \citenamefont {Ferioli}, \citenamefont {Masi}, \citenamefont
  {Mazzinghi}, \citenamefont {Wolswijk}, \citenamefont {Minardi}, \citenamefont
  {Modugno}, \citenamefont {Modugno}, \citenamefont {Inguscio},\ and\
  \citenamefont {Fattori}}]{Semeghini2018}%
  \BibitemOpen
  \bibfield  {author} {\bibinfo {author} {\bibfnamefont {G.}~\bibnamefont
  {Semeghini}}, \bibinfo {author} {\bibfnamefont {G.}~\bibnamefont {Ferioli}},
  \bibinfo {author} {\bibfnamefont {L.}~\bibnamefont {Masi}}, \bibinfo {author}
  {\bibfnamefont {C.}~\bibnamefont {Mazzinghi}}, \bibinfo {author}
  {\bibfnamefont {L.}~\bibnamefont {Wolswijk}}, \bibinfo {author}
  {\bibfnamefont {F.}~\bibnamefont {Minardi}}, \bibinfo {author} {\bibfnamefont
  {M.}~\bibnamefont {Modugno}}, \bibinfo {author} {\bibfnamefont
  {G.}~\bibnamefont {Modugno}}, \bibinfo {author} {\bibfnamefont
  {M.}~\bibnamefont {Inguscio}}, \ and\ \bibinfo {author} {\bibfnamefont
  {M.}~\bibnamefont {Fattori}},\ }\href {\doibase
  10.1103/PhysRevLett.120.235301} {\bibfield  {journal} {\bibinfo  {journal}
  {Phys. Rev. Lett.}\ }\textbf {\bibinfo {volume} {120}},\ \bibinfo {pages}
  {235301} (\bibinfo {year} {2018})}\BibitemShut {NoStop}%
\bibitem [{\citenamefont {Cheiney}\ \emph {et~al.}(2018)\citenamefont
  {Cheiney}, \citenamefont {Cabrera}, \citenamefont {Sanz}, \citenamefont
  {Naylor}, \citenamefont {Tanzi},\ and\ \citenamefont
  {Tarruell}}]{Cheiney2018}%
  \BibitemOpen
  \bibfield  {author} {\bibinfo {author} {\bibfnamefont {P.}~\bibnamefont
  {Cheiney}}, \bibinfo {author} {\bibfnamefont {C.~R.}\ \bibnamefont
  {Cabrera}}, \bibinfo {author} {\bibfnamefont {J.}~\bibnamefont {Sanz}},
  \bibinfo {author} {\bibfnamefont {B.}~\bibnamefont {Naylor}}, \bibinfo
  {author} {\bibfnamefont {L.}~\bibnamefont {Tanzi}}, \ and\ \bibinfo {author}
  {\bibfnamefont {L.}~\bibnamefont {Tarruell}},\ }\href {\doibase
  10.1103/PhysRevLett.120.135301} {\bibfield  {journal} {\bibinfo  {journal}
  {Phys. Rev. Lett.}\ }\textbf {\bibinfo {volume} {120}},\ \bibinfo {pages}
  {135301} (\bibinfo {year} {2018})}\BibitemShut {NoStop}%
\bibitem [{\citenamefont {D'Errico}\ \emph {et~al.}(2019)\citenamefont
  {D'Errico}, \citenamefont {Burchianti}, \citenamefont {Prevedelli},
  \citenamefont {Salasnich}, \citenamefont {Ancilotto}, \citenamefont
  {Modugno}, \citenamefont {Minardi},\ and\ \citenamefont {Fort}}]{Errico2019}%
  \BibitemOpen
  \bibfield  {author} {\bibinfo {author} {\bibfnamefont {C.}~\bibnamefont
  {D'Errico}}, \bibinfo {author} {\bibfnamefont {A.}~\bibnamefont
  {Burchianti}}, \bibinfo {author} {\bibfnamefont {M.}~\bibnamefont
  {Prevedelli}}, \bibinfo {author} {\bibfnamefont {L.}~\bibnamefont
  {Salasnich}}, \bibinfo {author} {\bibfnamefont {F.}~\bibnamefont
  {Ancilotto}}, \bibinfo {author} {\bibfnamefont {M.}~\bibnamefont {Modugno}},
  \bibinfo {author} {\bibfnamefont {F.}~\bibnamefont {Minardi}}, \ and\
  \bibinfo {author} {\bibfnamefont {C.}~\bibnamefont {Fort}},\ }\href {\doibase
  10.1103/PhysRevResearch.1.033155} {\bibfield  {journal} {\bibinfo  {journal}
  {Phys. Rev. Research}\ }\textbf {\bibinfo {volume} {1}},\ \bibinfo {pages}
  {033155} (\bibinfo {year} {2019})}\BibitemShut {NoStop}%
\bibitem [{\citenamefont {Guo}\ \emph {et~al.}(2021)\citenamefont {Guo},
  \citenamefont {Jia}, \citenamefont {Li}, \citenamefont {Ma}, \citenamefont
  {Hutson}, \citenamefont {Cui},\ and\ \citenamefont {Wang}}]{ZGuo2021}%
  \BibitemOpen
  \bibfield  {author} {\bibinfo {author} {\bibfnamefont {Z.}~\bibnamefont
  {Guo}}, \bibinfo {author} {\bibfnamefont {F.}~\bibnamefont {Jia}}, \bibinfo
  {author} {\bibfnamefont {L.}~\bibnamefont {Li}}, \bibinfo {author}
  {\bibfnamefont {Y.}~\bibnamefont {Ma}}, \bibinfo {author} {\bibfnamefont
  {J.~M.}\ \bibnamefont {Hutson}}, \bibinfo {author} {\bibfnamefont
  {X.}~\bibnamefont {Cui}}, \ and\ \bibinfo {author} {\bibfnamefont
  {D.}~\bibnamefont {Wang}},\ }\href {\doibase
  10.1103/PhysRevResearch.3.033247} {\bibfield  {journal} {\bibinfo  {journal}
  {Phys. Rev. Research}\ }\textbf {\bibinfo {volume} {3}},\ \bibinfo {pages}
  {033247} (\bibinfo {year} {2021})}\BibitemShut {NoStop}%
\bibitem [{\citenamefont {Edler}\ \emph {et~al.}(2017)\citenamefont {Edler},
  \citenamefont {Mishra}, \citenamefont {W\"achtler}, \citenamefont {Nath},
  \citenamefont {Sinha},\ and\ \citenamefont {Santos}}]{Edler2017}%
  \BibitemOpen
  \bibfield  {author} {\bibinfo {author} {\bibfnamefont {D.}~\bibnamefont
  {Edler}}, \bibinfo {author} {\bibfnamefont {C.}~\bibnamefont {Mishra}},
  \bibinfo {author} {\bibfnamefont {F.}~\bibnamefont {W\"achtler}}, \bibinfo
  {author} {\bibfnamefont {R.}~\bibnamefont {Nath}}, \bibinfo {author}
  {\bibfnamefont {S.}~\bibnamefont {Sinha}}, \ and\ \bibinfo {author}
  {\bibfnamefont {L.}~\bibnamefont {Santos}},\ }\href {\doibase
  10.1103/PhysRevLett.119.050403} {\bibfield  {journal} {\bibinfo  {journal}
  {Phys. Rev. Lett.}\ }\textbf {\bibinfo {volume} {119}},\ \bibinfo {pages}
  {050403} (\bibinfo {year} {2017})}\BibitemShut {NoStop}%
\bibitem [{\citenamefont {Salasnich}(2018)}]{Salasnich2018}%
  \BibitemOpen
  \bibfield  {author} {\bibinfo {author} {\bibfnamefont {L.}~\bibnamefont
  {Salasnich}},\ }\href {https://www.mdpi.com/2076-3417/8/10/1998} {\bibfield
  {journal} {\bibinfo  {journal} {Applied Sciences}\ }\textbf {\bibinfo
  {volume} {8}} (\bibinfo {year} {2018})}\BibitemShut {NoStop}%
\bibitem [{\citenamefont {Cidrim}\ \emph {et~al.}(2018)\citenamefont {Cidrim},
  \citenamefont {dos Santos}, \citenamefont {Henn},\ and\ \citenamefont
  {Macr\`{\i}}}]{Cidrim2018}%
  \BibitemOpen
  \bibfield  {author} {\bibinfo {author} {\bibfnamefont {A.}~\bibnamefont
  {Cidrim}}, \bibinfo {author} {\bibfnamefont {F.~E.~A.}\ \bibnamefont {dos
  Santos}}, \bibinfo {author} {\bibfnamefont {E.~A.~L.}\ \bibnamefont {Henn}},
  \ and\ \bibinfo {author} {\bibfnamefont {T.}~\bibnamefont {Macr\`{\i}}},\
  }\href {\doibase 10.1103/PhysRevA.98.023618} {\bibfield  {journal} {\bibinfo
  {journal} {Phys. Rev. A}\ }\textbf {\bibinfo {volume} {98}},\ \bibinfo
  {pages} {023618} (\bibinfo {year} {2018})}\BibitemShut {NoStop}%
\bibitem [{\citenamefont {Roccuzzo}\ and\ \citenamefont
  {Ancilotto}(2019)}]{Roccuzzo2019}%
  \BibitemOpen
  \bibfield  {author} {\bibinfo {author} {\bibfnamefont {S.~M.}\ \bibnamefont
  {Roccuzzo}}\ and\ \bibinfo {author} {\bibfnamefont {F.}~\bibnamefont
  {Ancilotto}},\ }\href {\doibase 10.1103/PhysRevA.99.041601} {\bibfield
  {journal} {\bibinfo  {journal} {Phys. Rev. A}\ }\textbf {\bibinfo {volume}
  {99}},\ \bibinfo {pages} {041601} (\bibinfo {year} {2019})}\BibitemShut
  {NoStop}%
\bibitem [{\citenamefont {Blakie}\ \emph {et~al.}(2020)\citenamefont {Blakie},
  \citenamefont {Baillie}, \citenamefont {Chomaz},\ and\ \citenamefont
  {Ferlaino}}]{Blakie2020}%
  \BibitemOpen
  \bibfield  {author} {\bibinfo {author} {\bibfnamefont {P.~B.}\ \bibnamefont
  {Blakie}}, \bibinfo {author} {\bibfnamefont {D.}~\bibnamefont {Baillie}},
  \bibinfo {author} {\bibfnamefont {L.}~\bibnamefont {Chomaz}}, \ and\ \bibinfo
  {author} {\bibfnamefont {F.}~\bibnamefont {Ferlaino}},\ }\href {\doibase
  10.1103/PhysRevResearch.2.043318} {\bibfield  {journal} {\bibinfo  {journal}
  {Phys. Rev. Research}\ }\textbf {\bibinfo {volume} {2}},\ \bibinfo {pages}
  {043318} (\bibinfo {year} {2020})}\BibitemShut {NoStop}%
\bibitem [{\citenamefont {Pal}\ \emph {et~al.}(2020)\citenamefont {Pal},
  \citenamefont {Baillie},\ and\ \citenamefont {Blakie}}]{Pal2020}%
  \BibitemOpen
  \bibfield  {author} {\bibinfo {author} {\bibfnamefont {S.}~\bibnamefont
  {Pal}}, \bibinfo {author} {\bibfnamefont {D.}~\bibnamefont {Baillie}}, \ and\
  \bibinfo {author} {\bibfnamefont {P.~B.}\ \bibnamefont {Blakie}},\ }\href
  {\doibase 10.1103/PhysRevA.102.043306} {\bibfield  {journal} {\bibinfo
  {journal} {Phys. Rev. A}\ }\textbf {\bibinfo {volume} {102}},\ \bibinfo
  {pages} {043306} (\bibinfo {year} {2020})}\BibitemShut {NoStop}%
\bibitem [{\citenamefont {Lee}\ \emph {et~al.}(2021)\citenamefont {Lee},
  \citenamefont {Baillie},\ and\ \citenamefont {Blakie}}]{Lee2021}%
  \BibitemOpen
  \bibfield  {author} {\bibinfo {author} {\bibfnamefont {A.-C.}\ \bibnamefont
  {Lee}}, \bibinfo {author} {\bibfnamefont {D.}~\bibnamefont {Baillie}}, \ and\
  \bibinfo {author} {\bibfnamefont {P.~B.}\ \bibnamefont {Blakie}},\ }\href
  {\doibase 10.1103/PhysRevResearch.3.013283} {\bibfield  {journal} {\bibinfo
  {journal} {Phys. Rev. Research}\ }\textbf {\bibinfo {volume} {3}},\ \bibinfo
  {pages} {013283} (\bibinfo {year} {2021})}\BibitemShut {NoStop}%
\bibitem [{\citenamefont {Böttcher}\ \emph {et~al.}(2021)\citenamefont
  {Böttcher}, \citenamefont {Schmidt}, \citenamefont {Hertkorn}, \citenamefont
  {Ng}, \citenamefont {Graham}, \citenamefont {Guo}, \citenamefont {Langen},\
  and\ \citenamefont {Pfau}}]{Bottcher2021}%
  \BibitemOpen
  \bibfield  {author} {\bibinfo {author} {\bibfnamefont {F.}~\bibnamefont
  {Böttcher}}, \bibinfo {author} {\bibfnamefont {J.-N.}\ \bibnamefont
  {Schmidt}}, \bibinfo {author} {\bibfnamefont {J.}~\bibnamefont {Hertkorn}},
  \bibinfo {author} {\bibfnamefont {K.~S.~H.}\ \bibnamefont {Ng}}, \bibinfo
  {author} {\bibfnamefont {S.~D.}\ \bibnamefont {Graham}}, \bibinfo {author}
  {\bibfnamefont {M.}~\bibnamefont {Guo}}, \bibinfo {author} {\bibfnamefont
  {T.}~\bibnamefont {Langen}}, \ and\ \bibinfo {author} {\bibfnamefont
  {T.}~\bibnamefont {Pfau}},\ }\href {\doibase 10.1088/1361-6633/abc9ab}
  {\bibfield  {journal} {\bibinfo  {journal} {Reports on Progress in Physics}\
  }\textbf {\bibinfo {volume} {84}},\ \bibinfo {pages} {012403} (\bibinfo
  {year} {2021})}\BibitemShut {NoStop}%
\bibitem [{\citenamefont {Luo}\ \emph {et~al.}(2021)\citenamefont {Luo},
  \citenamefont {Pang}, \citenamefont {Liu}, \citenamefont {Li},\ and\
  \citenamefont {Malomed}}]{Luo2021}%
  \BibitemOpen
  \bibfield  {author} {\bibinfo {author} {\bibfnamefont {Z.-H.}\ \bibnamefont
  {Luo}}, \bibinfo {author} {\bibfnamefont {W.}~\bibnamefont {Pang}}, \bibinfo
  {author} {\bibfnamefont {B.}~\bibnamefont {Liu}}, \bibinfo {author}
  {\bibfnamefont {Y.-Y.}\ \bibnamefont {Li}}, \ and\ \bibinfo {author}
  {\bibfnamefont {B.~A.}\ \bibnamefont {Malomed}},\ }\href {\doibase
  10.1007/s11467-020-1020-2} {\bibfield  {journal} {\bibinfo  {journal}
  {Frontiers of Physics}\ }\textbf {\bibinfo {volume} {16}},\ \bibinfo {pages}
  {32201} (\bibinfo {year} {2021})}\BibitemShut {NoStop}%
\bibitem [{\citenamefont {Cikojević}\ \emph {et~al.}(2020)\citenamefont
  {Cikojević}, \citenamefont {Markić},\ and\ \citenamefont
  {Boronat}}]{Cikojevic2020b}%
  \BibitemOpen
  \bibfield  {author} {\bibinfo {author} {\bibfnamefont {V.}~\bibnamefont
  {Cikojević}}, \bibinfo {author} {\bibfnamefont {L.~V.}\ \bibnamefont
  {Markić}}, \ and\ \bibinfo {author} {\bibfnamefont {J.}~\bibnamefont
  {Boronat}},\ }\href {\doibase 10.1088/1367-2630/ab867a} {\bibfield  {journal}
  {\bibinfo  {journal} {New Journal of Physics}\ }\textbf {\bibinfo {volume}
  {22}},\ \bibinfo {pages} {053045} (\bibinfo {year} {2020})}\BibitemShut
  {NoStop}%
\bibitem [{\citenamefont {Astrakharchik}\ and\ \citenamefont
  {Malomed}(2018)}]{Astrakharchik2018}%
  \BibitemOpen
  \bibfield  {author} {\bibinfo {author} {\bibfnamefont {G.~E.}\ \bibnamefont
  {Astrakharchik}}\ and\ \bibinfo {author} {\bibfnamefont {B.~A.}\ \bibnamefont
  {Malomed}},\ }\href {\doibase 10.1103/PhysRevA.98.013631} {\bibfield
  {journal} {\bibinfo  {journal} {Phys. Rev. A}\ }\textbf {\bibinfo {volume}
  {98}},\ \bibinfo {pages} {013631} (\bibinfo {year} {2018})}\BibitemShut
  {NoStop}%
\bibitem [{\citenamefont {Chiquillo}(2018{\natexlab{a}})}]{Chiquillo2018b}%
  \BibitemOpen
  \bibfield  {author} {\bibinfo {author} {\bibfnamefont {E.}~\bibnamefont
  {Chiquillo}},\ }\href {\doibase 10.1103/PhysRevA.97.063605} {\bibfield
  {journal} {\bibinfo  {journal} {Phys. Rev. A}\ }\textbf {\bibinfo {volume}
  {97}},\ \bibinfo {pages} {063605} (\bibinfo {year}
  {2018}{\natexlab{a}})}\BibitemShut {NoStop}%
\bibitem [{\citenamefont {Chiquillo}(2019)}]{Chiquillo2019}%
  \BibitemOpen
  \bibfield  {author} {\bibinfo {author} {\bibfnamefont {E.}~\bibnamefont
  {Chiquillo}},\ }\href {\doibase 10.1103/PhysRevA.99.051601} {\bibfield
  {journal} {\bibinfo  {journal} {Phys. Rev. A}\ }\textbf {\bibinfo {volume}
  {99}},\ \bibinfo {pages} {051601} (\bibinfo {year} {2019})}\BibitemShut
  {NoStop}%
\bibitem [{\citenamefont {Parisi}\ \emph {et~al.}(2019)\citenamefont {Parisi},
  \citenamefont {Astrakharchik},\ and\ \citenamefont {Giorgini}}]{Parisi2019}%
  \BibitemOpen
  \bibfield  {author} {\bibinfo {author} {\bibfnamefont {L.}~\bibnamefont
  {Parisi}}, \bibinfo {author} {\bibfnamefont {G.~E.}\ \bibnamefont
  {Astrakharchik}}, \ and\ \bibinfo {author} {\bibfnamefont {S.}~\bibnamefont
  {Giorgini}},\ }\href {\doibase 10.1103/PhysRevLett.122.105302} {\bibfield
  {journal} {\bibinfo  {journal} {Phys. Rev. Lett.}\ }\textbf {\bibinfo
  {volume} {122}},\ \bibinfo {pages} {105302} (\bibinfo {year}
  {2019})}\BibitemShut {NoStop}%
\bibitem [{\citenamefont {De~Rosi}\ \emph {et~al.}(2021)\citenamefont
  {De~Rosi}, \citenamefont {Astrakharchik},\ and\ \citenamefont
  {Massignan}}]{DeRosi2021}%
  \BibitemOpen
  \bibfield  {author} {\bibinfo {author} {\bibfnamefont {G.}~\bibnamefont
  {De~Rosi}}, \bibinfo {author} {\bibfnamefont {G.~E.}\ \bibnamefont
  {Astrakharchik}}, \ and\ \bibinfo {author} {\bibfnamefont {P.}~\bibnamefont
  {Massignan}},\ }\href {\doibase 10.1103/PhysRevA.103.043316} {\bibfield
  {journal} {\bibinfo  {journal} {Phys. Rev. A}\ }\textbf {\bibinfo {volume}
  {103}},\ \bibinfo {pages} {043316} (\bibinfo {year} {2021})}\BibitemShut
  {NoStop}%
\bibitem [{\citenamefont {Mistakidis}\ \emph {et~al.}(2021)\citenamefont
  {Mistakidis}, \citenamefont {Mithun}, \citenamefont {Kevrekidis},
  \citenamefont {Sadeghpour},\ and\ \citenamefont
  {Schmelcher}}]{Mistakidis2021}%
  \BibitemOpen
  \bibfield  {author} {\bibinfo {author} {\bibfnamefont {S.~I.}\ \bibnamefont
  {Mistakidis}}, \bibinfo {author} {\bibfnamefont {T.}~\bibnamefont {Mithun}},
  \bibinfo {author} {\bibfnamefont {P.~G.}\ \bibnamefont {Kevrekidis}},
  \bibinfo {author} {\bibfnamefont {H.~R.}\ \bibnamefont {Sadeghpour}}, \ and\
  \bibinfo {author} {\bibfnamefont {P.}~\bibnamefont {Schmelcher}},\ }\href
  {\doibase 10.1103/PhysRevResearch.3.043128} {\bibfield  {journal} {\bibinfo
  {journal} {Phys. Rev. Research}\ }\textbf {\bibinfo {volume} {3}},\ \bibinfo
  {pages} {043128} (\bibinfo {year} {2021})}\BibitemShut {NoStop}%
\bibitem [{\citenamefont {Zin}\ \emph {et~al.}(2018)\citenamefont {Zin},
  \citenamefont {Pylak}, \citenamefont {Wasak}, \citenamefont {Gajda},\ and\
  \citenamefont {Idziaszek}}]{Zin2018}%
  \BibitemOpen
  \bibfield  {author} {\bibinfo {author} {\bibfnamefont {P.}~\bibnamefont
  {Zin}}, \bibinfo {author} {\bibfnamefont {M.}~\bibnamefont {Pylak}}, \bibinfo
  {author} {\bibfnamefont {T.}~\bibnamefont {Wasak}}, \bibinfo {author}
  {\bibfnamefont {M.}~\bibnamefont {Gajda}}, \ and\ \bibinfo {author}
  {\bibfnamefont {Z.}~\bibnamefont {Idziaszek}},\ }\href {\doibase
  10.1103/PhysRevA.98.051603} {\bibfield  {journal} {\bibinfo  {journal} {Phys.
  Rev. A}\ }\textbf {\bibinfo {volume} {98}},\ \bibinfo {pages} {051603}
  (\bibinfo {year} {2018})}\BibitemShut {NoStop}%
\bibitem [{\citenamefont {Ilg}\ \emph {et~al.}(2018)\citenamefont {Ilg},
  \citenamefont {Kumlin}, \citenamefont {Santos}, \citenamefont {Petrov},\ and\
  \citenamefont {B\"uchler}}]{Ilg2018}%
  \BibitemOpen
  \bibfield  {author} {\bibinfo {author} {\bibfnamefont {T.}~\bibnamefont
  {Ilg}}, \bibinfo {author} {\bibfnamefont {J.}~\bibnamefont {Kumlin}},
  \bibinfo {author} {\bibfnamefont {L.}~\bibnamefont {Santos}}, \bibinfo
  {author} {\bibfnamefont {D.~S.}\ \bibnamefont {Petrov}}, \ and\ \bibinfo
  {author} {\bibfnamefont {H.~P.}\ \bibnamefont {B\"uchler}},\ }\href {\doibase
  10.1103/PhysRevA.98.051604} {\bibfield  {journal} {\bibinfo  {journal} {Phys.
  Rev. A}\ }\textbf {\bibinfo {volume} {98}},\ \bibinfo {pages} {051604}
  (\bibinfo {year} {2018})}\BibitemShut {NoStop}%
\bibitem [{\citenamefont {Lavoine}\ and\ \citenamefont
  {Bourdel}(2021)}]{Lavoine2021}%
  \BibitemOpen
  \bibfield  {author} {\bibinfo {author} {\bibfnamefont {L.}~\bibnamefont
  {Lavoine}}\ and\ \bibinfo {author} {\bibfnamefont {T.}~\bibnamefont
  {Bourdel}},\ }\href {\doibase 10.1103/PhysRevA.103.033312} {\bibfield
  {journal} {\bibinfo  {journal} {Phys. Rev. A}\ }\textbf {\bibinfo {volume}
  {103}},\ \bibinfo {pages} {033312} (\bibinfo {year} {2021})}\BibitemShut
  {NoStop}%
\bibitem [{\citenamefont {Cappellaro}\ \emph {et~al.}(2018)\citenamefont
  {Cappellaro}, \citenamefont {Macr\`{\i}},\ and\ \citenamefont
  {Salasnich}}]{Cappellaro2018}%
  \BibitemOpen
  \bibfield  {author} {\bibinfo {author} {\bibfnamefont {A.}~\bibnamefont
  {Cappellaro}}, \bibinfo {author} {\bibfnamefont {T.}~\bibnamefont
  {Macr\`{\i}}}, \ and\ \bibinfo {author} {\bibfnamefont {L.}~\bibnamefont
  {Salasnich}},\ }\href {\doibase 10.1103/PhysRevA.97.053623} {\bibfield
  {journal} {\bibinfo  {journal} {Phys. Rev. A}\ }\textbf {\bibinfo {volume}
  {97}},\ \bibinfo {pages} {053623} (\bibinfo {year} {2018})}\BibitemShut
  {NoStop}%
\bibitem [{\citenamefont {Tylutki}\ \emph {et~al.}(2020)\citenamefont
  {Tylutki}, \citenamefont {Astrakharchik}, \citenamefont {Malomed},\ and\
  \citenamefont {Petrov}}]{Tylutki2020}%
  \BibitemOpen
  \bibfield  {author} {\bibinfo {author} {\bibfnamefont {M.}~\bibnamefont
  {Tylutki}}, \bibinfo {author} {\bibfnamefont {G.~E.}\ \bibnamefont
  {Astrakharchik}}, \bibinfo {author} {\bibfnamefont {B.~A.}\ \bibnamefont
  {Malomed}}, \ and\ \bibinfo {author} {\bibfnamefont {D.~S.}\ \bibnamefont
  {Petrov}},\ }\href {\doibase 10.1103/PhysRevA.101.051601} {\bibfield
  {journal} {\bibinfo  {journal} {Phys. Rev. A}\ }\textbf {\bibinfo {volume}
  {101}},\ \bibinfo {pages} {051601} (\bibinfo {year} {2020})}\BibitemShut
  {NoStop}%
\bibitem [{\citenamefont {Tengstrand}\ and\ \citenamefont
  {Reimann}(2022)}]{NilssonTengstrand2022}%
  \BibitemOpen
  \bibfield  {author} {\bibinfo {author} {\bibfnamefont {M.~N.}\ \bibnamefont
  {Tengstrand}}\ and\ \bibinfo {author} {\bibfnamefont {S.~M.}\ \bibnamefont
  {Reimann}},\ }\href {\doibase 10.1103/PhysRevA.105.033319} {\bibfield
  {journal} {\bibinfo  {journal} {Phys. Rev. A}\ }\textbf {\bibinfo {volume}
  {105}},\ \bibinfo {pages} {033319} (\bibinfo {year} {2022})}\BibitemShut
  {NoStop}%
\bibitem [{\citenamefont {Wang}\ \emph {et~al.}(2021)\citenamefont {Wang},
  \citenamefont {Liu},\ and\ \citenamefont {Hu}}]{Wang2021}%
  \BibitemOpen
  \bibfield  {author} {\bibinfo {author} {\bibfnamefont {J.}~\bibnamefont
  {Wang}}, \bibinfo {author} {\bibfnamefont {X.-J.}\ \bibnamefont {Liu}}, \
  and\ \bibinfo {author} {\bibfnamefont {H.}~\bibnamefont {Hu}},\ }\href
  {\doibase 10.1088/1674-1056/abd2ad} {\bibfield  {journal} {\bibinfo
  {journal} {Chinese Physics B}\ }\textbf {\bibinfo {volume} {30}},\ \bibinfo
  {pages} {010306} (\bibinfo {year} {2021})}\BibitemShut {NoStop}%
\bibitem [{\citenamefont {Ota}\ and\ \citenamefont
  {Astrakharchik}(2020)}]{Ota2020a}%
  \BibitemOpen
  \bibfield  {author} {\bibinfo {author} {\bibfnamefont {M.}~\bibnamefont
  {Ota}}\ and\ \bibinfo {author} {\bibfnamefont {G.~E.}\ \bibnamefont
  {Astrakharchik}},\ }\href {\doibase 10.21468/SciPostPhys.9.2.020} {\bibfield
  {journal} {\bibinfo  {journal} {SciPost Phys.}\ }\textbf {\bibinfo {volume}
  {9}},\ \bibinfo {pages} {20} (\bibinfo {year} {2020})}\BibitemShut {NoStop}%
\bibitem [{\citenamefont {Cikojevi\ifmmode~\acute{c}\else \'{c}\fi{}}\ \emph
  {et~al.}(2020)\citenamefont {Cikojevi\ifmmode~\acute{c}\else \'{c}\fi{}},
  \citenamefont {Marki\ifmmode~\acute{c}\else \'{c}\fi{}}, \citenamefont {Pi},
  \citenamefont {Barranco},\ and\ \citenamefont {Boronat}}]{Cikojevic2020}%
  \BibitemOpen
  \bibfield  {author} {\bibinfo {author} {\bibfnamefont {V.}~\bibnamefont
  {Cikojevi\ifmmode~\acute{c}\else \'{c}\fi{}}}, \bibinfo {author}
  {\bibfnamefont {L.~V. c.~v.}\ \bibnamefont {Marki\ifmmode~\acute{c}\else
  \'{c}\fi{}}}, \bibinfo {author} {\bibfnamefont {M.}~\bibnamefont {Pi}},
  \bibinfo {author} {\bibfnamefont {M.}~\bibnamefont {Barranco}}, \ and\
  \bibinfo {author} {\bibfnamefont {J.}~\bibnamefont {Boronat}},\ }\href
  {\doibase 10.1103/PhysRevA.102.033335} {\bibfield  {journal} {\bibinfo
  {journal} {Phys. Rev. A}\ }\textbf {\bibinfo {volume} {102}},\ \bibinfo
  {pages} {033335} (\bibinfo {year} {2020})}\BibitemShut {NoStop}%
\bibitem [{\citenamefont {Kopyciński}\ \emph {et~al.}(2023)\citenamefont
  {Kopyciński}, \citenamefont {Parisi}, \citenamefont {Parker},\ and\
  \citenamefont {Pawłowski}}]{Kopycinski2023}%
  \BibitemOpen
  \bibfield  {author} {\bibinfo {author} {\bibfnamefont {J.}~\bibnamefont
  {Kopyciński}}, \bibinfo {author} {\bibfnamefont {L.}~\bibnamefont {Parisi}},
  \bibinfo {author} {\bibfnamefont {N.~G.}\ \bibnamefont {Parker}}, \ and\
  \bibinfo {author} {\bibfnamefont {K.}~\bibnamefont {Pawłowski}},\ } {\  \href
  {\doibase 10.48550/ARXIV.2301.04417} {arXiv:2301.04417}} (\bibinfo {year} {2023})\BibitemShut {NoStop}%
\bibitem [{\citenamefont {Lieb}\ and\ \citenamefont
  {Liniger}(1963)}]{LiebLiniger1963}%
  \BibitemOpen
  \bibfield  {author} {\bibinfo {author} {\bibfnamefont {E.~H.}\ \bibnamefont
  {Lieb}}\ and\ \bibinfo {author} {\bibfnamefont {W.}~\bibnamefont {Liniger}},\
  }\href {\doibase 10.1103/PhysRev.130.1605} {\bibfield  {journal} {\bibinfo
  {journal} {Phys. Rev.}\ }\textbf {\bibinfo {volume} {130}},\ \bibinfo {pages}
  {1605} (\bibinfo {year} {1963})}\BibitemShut {NoStop}%
\bibitem [{\citenamefont {Lieb}(1963)}]{Lieb1963}%
  \BibitemOpen
  \bibfield  {author} {\bibinfo {author} {\bibfnamefont {E.~H.}\ \bibnamefont
  {Lieb}},\ }\href {\doibase 10.1103/PhysRev.130.1616} {\bibfield  {journal}
  {\bibinfo  {journal} {Phys. Rev.}\ }\textbf {\bibinfo {volume} {130}},\
  \bibinfo {pages} {1616} (\bibinfo {year} {1963})}\BibitemShut {NoStop}%
\bibitem [{\citenamefont {Parisi}\ and\ \citenamefont
  {Giorgini}(2020)}]{Parisi2020}%
  \BibitemOpen
  \bibfield  {author} {\bibinfo {author} {\bibfnamefont {L.}~\bibnamefont
  {Parisi}}\ and\ \bibinfo {author} {\bibfnamefont {S.}~\bibnamefont
  {Giorgini}},\ }\href {\doibase 10.1103/PhysRevA.102.023318} {\bibfield
  {journal} {\bibinfo  {journal} {Phys. Rev. A}\ }\textbf {\bibinfo {volume}
  {102}},\ \bibinfo {pages} {023318} (\bibinfo {year} {2020})}\BibitemShut
  {NoStop}%
\bibitem [{\citenamefont {Morera}\ \emph {et~al.}(2020)\citenamefont {Morera},
  \citenamefont {Astrakharchik}, \citenamefont {Polls},\ and\ \citenamefont
  {Juli\'a-D\'{\i}az}}]{Morera2020}%
  \BibitemOpen
  \bibfield  {author} {\bibinfo {author} {\bibfnamefont {I.}~\bibnamefont
  {Morera}}, \bibinfo {author} {\bibfnamefont {G.~E.}\ \bibnamefont
  {Astrakharchik}}, \bibinfo {author} {\bibfnamefont {A.}~\bibnamefont
  {Polls}}, \ and\ \bibinfo {author} {\bibfnamefont {B.}~\bibnamefont
  {Juli\'a-D\'{\i}az}},\ }\href {\doibase 10.1103/PhysRevResearch.2.022008}
  {\bibfield  {journal} {\bibinfo  {journal} {Phys. Rev. Research}\ }\textbf
  {\bibinfo {volume} {2}},\ \bibinfo {pages} {022008} (\bibinfo {year}
  {2020})}\BibitemShut {NoStop}%
\bibitem [{\citenamefont {Morera}\ \emph {et~al.}(2022)\citenamefont {Morera},
  \citenamefont {Juli\'a-D\'{\i}az},\ and\ \citenamefont
  {Valiente}}]{Morera2021a}%
  \BibitemOpen
  \bibfield  {author} {\bibinfo {author} {\bibfnamefont {I.}~\bibnamefont
  {Morera}}, \bibinfo {author} {\bibfnamefont {B.}~\bibnamefont
  {Juli\'a-D\'{\i}az}}, \ and\ \bibinfo {author} {\bibfnamefont
  {M.}~\bibnamefont {Valiente}},\ }\href {\doibase
  10.1103/PhysRevResearch.4.L042024} {\bibfield  {journal} {\bibinfo  {journal}
  {Phys. Rev. Res.}\ }\textbf {\bibinfo {volume} {4}},\ \bibinfo {pages}
  {L042024} (\bibinfo {year} {2022})}\BibitemShut {NoStop}%
\bibitem [{\citenamefont {Chiquillo}(2018{\natexlab{b}})}]{Chiquillo2018}%
  \BibitemOpen
  \bibfield  {author} {\bibinfo {author} {\bibfnamefont {E.}~\bibnamefont
  {Chiquillo}},\ }\href {\doibase 10.1103/PhysRevA.97.013614} {\bibfield
  {journal} {\bibinfo  {journal} {Phys. Rev. A}\ }\textbf {\bibinfo {volume}
  {97}},\ \bibinfo {pages} {013614} (\bibinfo {year}
  {2018}{\natexlab{b}})}\BibitemShut {NoStop}%
\bibitem [{\citenamefont {Serwane}\ \emph {et~al.}(2011)\citenamefont
  {Serwane}, \citenamefont {Z{\"u}rn}, \citenamefont {Lompe}, \citenamefont
  {Ottenstein}, \citenamefont {Wenz},\ and\ \citenamefont
  {Jochim}}]{Serwane2011}%
  \BibitemOpen
  \bibfield  {author} {\bibinfo {author} {\bibfnamefont {F.}~\bibnamefont
  {Serwane}}, \bibinfo {author} {\bibfnamefont {G.}~\bibnamefont {Z{\"u}rn}},
  \bibinfo {author} {\bibfnamefont {T.}~\bibnamefont {Lompe}}, \bibinfo
  {author} {\bibfnamefont {T.}~\bibnamefont {Ottenstein}}, \bibinfo {author}
  {\bibfnamefont {A.}~\bibnamefont {Wenz}}, \ and\ \bibinfo {author}
  {\bibfnamefont {S.}~\bibnamefont {Jochim}},\ }\href {\doibase
  10.1126/science.1201351} {\bibfield  {journal} {\bibinfo  {journal}
  {Science}\ }\textbf {\bibinfo {volume} {332}},\ \bibinfo {pages} {336}
  (\bibinfo {year} {2011})}\BibitemShut {NoStop}%
\bibitem [{\citenamefont {Z\"urn}\ \emph {et~al.}(2012)\citenamefont {Z\"urn},
  \citenamefont {Serwane}, \citenamefont {Lompe}, \citenamefont {Wenz},
  \citenamefont {Ries}, \citenamefont {Bohn},\ and\ \citenamefont
  {Jochim}}]{Zurn2012}%
  \BibitemOpen
  \bibfield  {author} {\bibinfo {author} {\bibfnamefont {G.}~\bibnamefont
  {Z\"urn}}, \bibinfo {author} {\bibfnamefont {F.}~\bibnamefont {Serwane}},
  \bibinfo {author} {\bibfnamefont {T.}~\bibnamefont {Lompe}}, \bibinfo
  {author} {\bibfnamefont {A.~N.}\ \bibnamefont {Wenz}}, \bibinfo {author}
  {\bibfnamefont {M.~G.}\ \bibnamefont {Ries}}, \bibinfo {author}
  {\bibfnamefont {J.~E.}\ \bibnamefont {Bohn}}, \ and\ \bibinfo {author}
  {\bibfnamefont {S.}~\bibnamefont {Jochim}},\ }\href {\doibase
  10.1103/PhysRevLett.108.075303} {\bibfield  {journal} {\bibinfo  {journal}
  {Phys. Rev. Lett.}\ }\textbf {\bibinfo {volume} {108}},\ \bibinfo {pages}
  {075303} (\bibinfo {year} {2012})}\BibitemShut {NoStop}%
\bibitem [{\citenamefont {Z\"urn}\ \emph {et~al.}(2013)\citenamefont {Z\"urn},
  \citenamefont {Wenz}, \citenamefont {Murmann}, \citenamefont {Bergschneider},
  \citenamefont {Lompe},\ and\ \citenamefont {Jochim}}]{Zurn2013}%
  \BibitemOpen
  \bibfield  {author} {\bibinfo {author} {\bibfnamefont {G.}~\bibnamefont
  {Z\"urn}}, \bibinfo {author} {\bibfnamefont {A.~N.}\ \bibnamefont {Wenz}},
  \bibinfo {author} {\bibfnamefont {S.}~\bibnamefont {Murmann}}, \bibinfo
  {author} {\bibfnamefont {A.}~\bibnamefont {Bergschneider}}, \bibinfo {author}
  {\bibfnamefont {T.}~\bibnamefont {Lompe}}, \ and\ \bibinfo {author}
  {\bibfnamefont {S.}~\bibnamefont {Jochim}},\ }\href {\doibase
  10.1103/PhysRevLett.111.175302} {\bibfield  {journal} {\bibinfo  {journal}
  {Phys. Rev. Lett.}\ }\textbf {\bibinfo {volume} {111}},\ \bibinfo {pages}
  {175302} (\bibinfo {year} {2013})}\BibitemShut {NoStop}%
\bibitem [{\citenamefont {Rontani}(2012)}]{Rontani2012}%
  \BibitemOpen
  \bibfield  {author} {\bibinfo {author} {\bibfnamefont {M.}~\bibnamefont
  {Rontani}},\ }\href {\doibase 10.1103/PhysRevLett.108.115302} {\bibfield
  {journal} {\bibinfo  {journal} {Phys. Rev. Lett.}\ }\textbf {\bibinfo
  {volume} {108}},\ \bibinfo {pages} {115302} (\bibinfo {year}
  {2012})}\BibitemShut {NoStop}%
\bibitem [{\citenamefont {Murmann}\ \emph {et~al.}(2015)\citenamefont
  {Murmann}, \citenamefont {Deuretzbacher}, \citenamefont {Z{\"u}rn},
  \citenamefont {Bjerlin}, \citenamefont {Reimann}, \citenamefont {Santos},
  \citenamefont {Lompe},\ and\ \citenamefont {Jochim}}]{Murmann2015}%
  \BibitemOpen
  \bibfield  {author} {\bibinfo {author} {\bibfnamefont {S.}~\bibnamefont
  {Murmann}}, \bibinfo {author} {\bibfnamefont {F.}~\bibnamefont
  {Deuretzbacher}}, \bibinfo {author} {\bibfnamefont {G.}~\bibnamefont
  {Z{\"u}rn}}, \bibinfo {author} {\bibfnamefont {J.}~\bibnamefont {Bjerlin}},
  \bibinfo {author} {\bibfnamefont {S.~M.}\ \bibnamefont {Reimann}}, \bibinfo
  {author} {\bibfnamefont {L.}~\bibnamefont {Santos}}, \bibinfo {author}
  {\bibfnamefont {T.}~\bibnamefont {Lompe}}, \ and\ \bibinfo {author}
  {\bibfnamefont {S.}~\bibnamefont {Jochim}},\ }\href {\doibase
  10.1103/PhysRevLett.115.215301} {\bibfield  {journal} {\bibinfo  {journal}
  {Phys. Rev. Lett.}\ }\textbf {\bibinfo {volume} {115}},\ \bibinfo {pages}
  {215301} (\bibinfo {year} {2015})}\BibitemShut {NoStop}%
\bibitem [{\citenamefont {Deuretzbacher}\ \emph {et~al.}(2014)\citenamefont
  {Deuretzbacher}, \citenamefont {Becker}, \citenamefont {Bjerlin},
  \citenamefont {Reimann},\ and\ \citenamefont {Santos}}]{Deuretzbacher2014}%
  \BibitemOpen
  \bibfield  {author} {\bibinfo {author} {\bibfnamefont {F.}~\bibnamefont
  {Deuretzbacher}}, \bibinfo {author} {\bibfnamefont {D.}~\bibnamefont
  {Becker}}, \bibinfo {author} {\bibfnamefont {J.}~\bibnamefont {Bjerlin}},
  \bibinfo {author} {\bibfnamefont {S.}~\bibnamefont {Reimann}}, \ and\
  \bibinfo {author} {\bibfnamefont {L.}~\bibnamefont {Santos}},\ }\href
  {\doibase 10.1103/PhysRevA.90.013611} {\bibfield  {journal} {\bibinfo
  {journal} {Phys. Rev. A}\ }\textbf {\bibinfo {volume} {90}},\ \bibinfo
  {pages} {013611} (\bibinfo {year} {2014})}\BibitemShut {NoStop}%
\bibitem [{\citenamefont {Wenz}\ \emph {et~al.}(2013)\citenamefont {Wenz},
  \citenamefont {Z{\"u}rn}, \citenamefont {Murmann}, \citenamefont {Brouzos},
  \citenamefont {Lompe},\ and\ \citenamefont {Jochim}}]{Wenz2013}%
  \BibitemOpen
  \bibfield  {author} {\bibinfo {author} {\bibfnamefont {A.}~\bibnamefont
  {Wenz}}, \bibinfo {author} {\bibfnamefont {G.}~\bibnamefont {Z{\"u}rn}},
  \bibinfo {author} {\bibfnamefont {S.}~\bibnamefont {Murmann}}, \bibinfo
  {author} {\bibfnamefont {I.}~\bibnamefont {Brouzos}}, \bibinfo {author}
  {\bibfnamefont {T.}~\bibnamefont {Lompe}}, \ and\ \bibinfo {author}
  {\bibfnamefont {S.}~\bibnamefont {Jochim}},\ }\href {\doibase
  10.1126/science.1240516} {\bibfield  {journal} {\bibinfo  {journal}
  {Science}\ }\textbf {\bibinfo {volume} {342}},\ \bibinfo {pages} {457}
  (\bibinfo {year} {2013})}\BibitemShut {NoStop}%
\bibitem [{\citenamefont {Bjerlin}\ \emph {et~al.}(2016)\citenamefont
  {Bjerlin}, \citenamefont {Reimann},\ and\ \citenamefont
  {Bruun}}]{Bjerlin2016}%
  \BibitemOpen
  \bibfield  {author} {\bibinfo {author} {\bibfnamefont {J.}~\bibnamefont
  {Bjerlin}}, \bibinfo {author} {\bibfnamefont {S.}~\bibnamefont {Reimann}}, \
  and\ \bibinfo {author} {\bibfnamefont {G.}~\bibnamefont {Bruun}},\ }\href
  {\doibase 10.1103/PhysRevLett.116.155302} {\bibfield  {journal} {\bibinfo
  {journal} {Phys. Rev. Lett.}\ }\textbf {\bibinfo {volume} {116}},\ \bibinfo
  {pages} {155302} (\bibinfo {year} {2016})}\BibitemShut {NoStop}%
\bibitem [{\citenamefont {Resare}\ and\ \citenamefont
  {Hofmann}(2022)}]{Resare2022}%
  \BibitemOpen
  \bibfield  {author} {\bibinfo {author} {\bibfnamefont {F.}~\bibnamefont
  {Resare}}\ and\ \bibinfo {author} {\bibfnamefont {J.}~\bibnamefont
  {Hofmann}},\ }\href {\doibase 10.48550/ARXIV.2208.03762} {arXiv:2208.03762}
   (\bibinfo {year} {2022})\BibitemShut {NoStop}%
\bibitem [{\citenamefont {Higgs}(1964)}]{Higgs1964}%
  \BibitemOpen
  \bibfield  {author} {\bibinfo {author} {\bibfnamefont {P.~W.}\ \bibnamefont
  {Higgs}},\ }\href {\doibase 10.1103/PhysRevLett.13.508} {\bibfield  {journal}
  {\bibinfo  {journal} {Phys. Rev. Lett.}\ }\textbf {\bibinfo {volume} {13}},\
  \bibinfo {pages} {508} (\bibinfo {year} {1964})}\BibitemShut {NoStop}%
\bibitem [{\citenamefont {Anderson}(1958)}]{Anderson1958}%
  \BibitemOpen
  \bibfield  {author} {\bibinfo {author} {\bibfnamefont {P.~W.}\ \bibnamefont
  {Anderson}},\ }\href {\doibase 10.1103/PhysRev.112.1900} {\bibfield
  {journal} {\bibinfo  {journal} {Phys. Rev.}\ }\textbf {\bibinfo {volume}
  {112}},\ \bibinfo {pages} {1900} (\bibinfo {year} {1958})}\BibitemShut
  {NoStop}%
\bibitem [{\citenamefont {Bruun}(2014)}]{Bruun2014}%
  \BibitemOpen
  \bibfield  {author} {\bibinfo {author} {\bibfnamefont {G.~M.}\ \bibnamefont
  {Bruun}},\ }\href {\doibase 10.1103/PhysRevA.90.023621} {\bibfield  {journal}
  {\bibinfo  {journal} {Phys. Rev. A}\ }\textbf {\bibinfo {volume} {90}},\
  \bibinfo {pages} {023621} (\bibinfo {year} {2014})}\BibitemShut {NoStop}%
\bibitem [{\citenamefont {Bayha}\ \emph {et~al.}(2020)\citenamefont {Bayha},
  \citenamefont {Holten}, \citenamefont {Klemt}, \citenamefont {Subramanian},
  \citenamefont {Bjerlin}, \citenamefont {Reimann}, \citenamefont {Bruun},
  \citenamefont {Preiss},\ and\ \citenamefont {Jochim}}]{Bayha2020}%
  \BibitemOpen
  \bibfield  {author} {\bibinfo {author} {\bibfnamefont {L.}~\bibnamefont
  {Bayha}}, \bibinfo {author} {\bibfnamefont {M.}~\bibnamefont {Holten}},
  \bibinfo {author} {\bibfnamefont {R.}~\bibnamefont {Klemt}}, \bibinfo
  {author} {\bibfnamefont {K.}~\bibnamefont {Subramanian}}, \bibinfo {author}
  {\bibfnamefont {J.}~\bibnamefont {Bjerlin}}, \bibinfo {author} {\bibfnamefont
  {S.~M.}\ \bibnamefont {Reimann}}, \bibinfo {author} {\bibfnamefont {G.~M.}\
  \bibnamefont {Bruun}}, \bibinfo {author} {\bibfnamefont {P.~M.}\ \bibnamefont
  {Preiss}}, \ and\ \bibinfo {author} {\bibfnamefont {S.}~\bibnamefont
  {Jochim}},\ }\href {\doibase 10.1038/s41586-020-2936-y} {\bibfield  {journal}
  {\bibinfo  {journal} {Nature}\ }\textbf {\bibinfo {volume} {587}},\ \bibinfo
  {pages} {583} (\bibinfo {year} {2020})}\BibitemShut {NoStop}%
\bibitem [{\citenamefont {Holten}\ \emph {et~al.}(2022)\citenamefont {Holten},
  \citenamefont {Bayha}, \citenamefont {Subramanian}, \citenamefont
  {Brandstetter}, \citenamefont {Heintze}, \citenamefont {Lunt}, \citenamefont
  {Preiss},\ and\ \citenamefont {Jochim}}]{Holten2022}%
  \BibitemOpen
  \bibfield  {author} {\bibinfo {author} {\bibfnamefont {M.}~\bibnamefont
  {Holten}}, \bibinfo {author} {\bibfnamefont {L.}~\bibnamefont {Bayha}},
  \bibinfo {author} {\bibfnamefont {K.}~\bibnamefont {Subramanian}}, \bibinfo
  {author} {\bibfnamefont {S.}~\bibnamefont {Brandstetter}}, \bibinfo {author}
  {\bibfnamefont {C.}~\bibnamefont {Heintze}}, \bibinfo {author} {\bibfnamefont
  {P.}~\bibnamefont {Lunt}}, \bibinfo {author} {\bibfnamefont {P.~M.}\
  \bibnamefont {Preiss}}, \ and\ \bibinfo {author} {\bibfnamefont
  {S.}~\bibnamefont {Jochim}},\ }\href {\doibase 10.1038/s41586-022-04678-1}
  {\bibfield  {journal} {\bibinfo  {journal} {Nature}\ }\textbf {\bibinfo
  {volume} {606}},\ \bibinfo {pages} {287} (\bibinfo {year}
  {2022})}\BibitemShut {NoStop}%
\bibitem [{\citenamefont {Kanamoto}\ \emph
  {et~al.}(2003{\natexlab{a}})\citenamefont {Kanamoto}, \citenamefont {Saito},\
  and\ \citenamefont {Ueda}}]{kanamoto2003quantum}%
  \BibitemOpen
  \bibfield  {author} {\bibinfo {author} {\bibfnamefont {R.}~\bibnamefont
  {Kanamoto}}, \bibinfo {author} {\bibfnamefont {H.}~\bibnamefont {Saito}}, \
  and\ \bibinfo {author} {\bibfnamefont {M.}~\bibnamefont {Ueda}},\ }\href
  {\doibase 10.1103/PhysRevA.67.013608} {\bibfield  {journal} {\bibinfo
  {journal} {Phys. Rev. A}\ }\textbf {\bibinfo {volume} {67}},\ \bibinfo
  {pages} {013608} (\bibinfo {year} {2003}{\natexlab{a}})}\BibitemShut
  {NoStop}%
\bibitem [{\citenamefont {Kanamoto}\ \emph
  {et~al.}(2003{\natexlab{b}})\citenamefont {Kanamoto}, \citenamefont {Saito},\
  and\ \citenamefont {Ueda}}]{kanamoto2003stability}%
  \BibitemOpen
  \bibfield  {author} {\bibinfo {author} {\bibfnamefont {R.}~\bibnamefont
  {Kanamoto}}, \bibinfo {author} {\bibfnamefont {H.}~\bibnamefont {Saito}}, \
  and\ \bibinfo {author} {\bibfnamefont {M.}~\bibnamefont {Ueda}},\ }\href
  {\doibase 10.1103/PhysRevA.68.043619} {\bibfield  {journal} {\bibinfo
  {journal} {Phys. Rev. A}\ }\textbf {\bibinfo {volume} {68}},\ \bibinfo
  {pages} {043619} (\bibinfo {year} {2003}{\natexlab{b}})}\BibitemShut
  {NoStop}%
\bibitem [{\citenamefont {Kanamoto}\ \emph {et~al.}(2005)\citenamefont
  {Kanamoto}, \citenamefont {Saito},\ and\ \citenamefont
  {Ueda}}]{kanamoto2005symmetry}%
  \BibitemOpen
  \bibfield  {author} {\bibinfo {author} {\bibfnamefont {R.}~\bibnamefont
  {Kanamoto}}, \bibinfo {author} {\bibfnamefont {H.}~\bibnamefont {Saito}}, \
  and\ \bibinfo {author} {\bibfnamefont {M.}~\bibnamefont {Ueda}},\ }\href
  {\doibase 10.1103/PhysRevLett.94.090404} {\bibfield  {journal} {\bibinfo
  {journal} {Phys. Rev. Lett.}\ }\textbf {\bibinfo {volume} {94}},\ \bibinfo
  {pages} {090404} (\bibinfo {year} {2005})}\BibitemShut {NoStop}%
\bibitem [{\citenamefont {Roth}(2009)}]{Roth2009}%
  \BibitemOpen
  \bibfield  {author} {\bibinfo {author} {\bibfnamefont {R.}~\bibnamefont
  {Roth}},\ }\href {\doibase 10.1103/PhysRevC.79.064324} {\bibfield  {journal}
  {\bibinfo  {journal} {Phys. Rev. C}\ }\textbf {\bibinfo {volume} {79}},\
  \bibinfo {pages} {064324} (\bibinfo {year} {2009})}\BibitemShut {NoStop}%
\bibitem [{\citenamefont {Hertkorn}\ \emph {et~al.}(2019)\citenamefont
  {Hertkorn}, \citenamefont {B\"ottcher}, \citenamefont {Guo}, \citenamefont
  {Schmidt}, \citenamefont {Langen}, \citenamefont {B\"uchler},\ and\
  \citenamefont {Pfau}}]{hertkorn2019fate}%
  \BibitemOpen
  \bibfield  {author} {\bibinfo {author} {\bibfnamefont {J.}~\bibnamefont
  {Hertkorn}}, \bibinfo {author} {\bibfnamefont {F.}~\bibnamefont
  {B\"ottcher}}, \bibinfo {author} {\bibfnamefont {M.}~\bibnamefont {Guo}},
  \bibinfo {author} {\bibfnamefont {J.~N.}\ \bibnamefont {Schmidt}}, \bibinfo
  {author} {\bibfnamefont {T.}~\bibnamefont {Langen}}, \bibinfo {author}
  {\bibfnamefont {H.~P.}\ \bibnamefont {B\"uchler}}, \ and\ \bibinfo {author}
  {\bibfnamefont {T.}~\bibnamefont {Pfau}},\ }\href {\doibase
  10.1103/PhysRevLett.123.193002} {\bibfield  {journal} {\bibinfo  {journal}
  {Phys. Rev. Lett.}\ }\textbf {\bibinfo {volume} {123}},\ \bibinfo {pages}
  {193002} (\bibinfo {year} {2019})}\BibitemShut {NoStop}%
\bibitem [{\citenamefont {Bohr}\ and\ \citenamefont
  {Mottelson}(1998)}]{BohrMottelson}%
  \BibitemOpen
  \bibfield  {author} {\bibinfo {author} {\bibfnamefont {A.}~\bibnamefont
  {Bohr}}\ and\ \bibinfo {author} {\bibfnamefont {B.}~\bibnamefont
  {Mottelson}},\ }\href@noop {} {\emph {\bibinfo {title} {Nuclear
  Structure}}},\ \bibinfo {series} {Nuclear Structure}\ No.\ \bibinfo {number}
  {v. 1}\ (\bibinfo  {publisher} {World Scientific},\ \bibinfo {year}
  {1998})\BibitemShut {NoStop}%
\bibitem [{\citenamefont {Bloch}(1973)}]{Bloch1973}%
  \BibitemOpen
  \bibfield  {author} {\bibinfo {author} {\bibfnamefont {F.}~\bibnamefont
  {Bloch}},\ }\href {\doibase 10.1103/PhysRevA.7.2187} {\bibfield  {journal}
  {\bibinfo  {journal} {Phys. Rev. A}\ }\textbf {\bibinfo {volume} {7}},\
  \bibinfo {pages} {2187} (\bibinfo {year} {1973})}\BibitemShut {NoStop}%
\bibitem [{\citenamefont {Smyrnakis}\ \emph {et~al.}(2009)\citenamefont
  {Smyrnakis}, \citenamefont {Bargi}, \citenamefont {Kavoulakis}, \citenamefont
  {Magiropoulos}, \citenamefont {K\"arkk\"ainen},\ and\ \citenamefont
  {Reimann}}]{Smyrnakis2009}%
  \BibitemOpen
  \bibfield  {author} {\bibinfo {author} {\bibfnamefont {J.}~\bibnamefont
  {Smyrnakis}}, \bibinfo {author} {\bibfnamefont {S.}~\bibnamefont {Bargi}},
  \bibinfo {author} {\bibfnamefont {G.~M.}\ \bibnamefont {Kavoulakis}},
  \bibinfo {author} {\bibfnamefont {M.}~\bibnamefont {Magiropoulos}}, \bibinfo
  {author} {\bibfnamefont {K.}~\bibnamefont {K\"arkk\"ainen}}, \ and\ \bibinfo
  {author} {\bibfnamefont {S.~M.}\ \bibnamefont {Reimann}},\ }\href {\doibase
  10.1103/PhysRevLett.103.100404} {\bibfield  {journal} {\bibinfo  {journal}
  {Phys. Rev. Lett.}\ }\textbf {\bibinfo {volume} {103}},\ \bibinfo {pages}
  {100404} (\bibinfo {year} {2009})}\BibitemShut {NoStop}%
\bibitem [{\citenamefont {Anoshkin}\ \emph {et~al.}(2013)\citenamefont
  {Anoshkin}, \citenamefont {Wu},\ and\ \citenamefont
  {Zaremba}}]{Anoshkin2013}%
  \BibitemOpen
  \bibfield  {author} {\bibinfo {author} {\bibfnamefont {K.}~\bibnamefont
  {Anoshkin}}, \bibinfo {author} {\bibfnamefont {Z.}~\bibnamefont {Wu}}, \ and\
  \bibinfo {author} {\bibfnamefont {E.}~\bibnamefont {Zaremba}},\ }\href
  {\doibase 10.1103/PhysRevA.88.013609} {\bibfield  {journal} {\bibinfo
  {journal} {Phys. Rev. A}\ }\textbf {\bibinfo {volume} {88}},\ \bibinfo
  {pages} {013609} (\bibinfo {year} {2013})}\BibitemShut {NoStop}%
\bibitem [{\citenamefont {Nilsson~Tengstrand}\ \emph
  {et~al.}(2021)\citenamefont {Nilsson~Tengstrand}, \citenamefont {Boholm},
  \citenamefont {Sachdeva}, \citenamefont {Bengtsson},\ and\ \citenamefont
  {Reimann}}]{NilssonTengstrand2021}%
  \BibitemOpen
  \bibfield  {author} {\bibinfo {author} {\bibfnamefont {M.}~\bibnamefont
  {Nilsson~Tengstrand}}, \bibinfo {author} {\bibfnamefont {D.}~\bibnamefont
  {Boholm}}, \bibinfo {author} {\bibfnamefont {R.}~\bibnamefont {Sachdeva}},
  \bibinfo {author} {\bibfnamefont {J.}~\bibnamefont {Bengtsson}}, \ and\
  \bibinfo {author} {\bibfnamefont {S.~M.}\ \bibnamefont {Reimann}},\ }\href
  {\doibase 10.1103/PhysRevA.103.013313} {\bibfield  {journal} {\bibinfo
  {journal} {Phys. Rev. A}\ }\textbf {\bibinfo {volume} {103}},\ \bibinfo
  {pages} {013313} (\bibinfo {year} {2021})}\BibitemShut {NoStop}%
\bibitem [{\citenamefont {\"Ogren}\ \emph {et~al.}(2021)\citenamefont
  {\"Ogren}, \citenamefont {Drougakis}, \citenamefont {Vasilakis},
  \citenamefont {von Klitzing},\ and\ \citenamefont {Kavoulakis}}]{Ogren2021}%
  \BibitemOpen
  \bibfield  {author} {\bibinfo {author} {\bibfnamefont {M.}~\bibnamefont
  {\"Ogren}}, \bibinfo {author} {\bibfnamefont {G.}~\bibnamefont {Drougakis}},
  \bibinfo {author} {\bibfnamefont {G.}~\bibnamefont {Vasilakis}}, \bibinfo
  {author} {\bibfnamefont {W.}~\bibnamefont {von Klitzing}}, \ and\ \bibinfo
  {author} {\bibfnamefont {G.~M.}\ \bibnamefont {Kavoulakis}},\ }\href
  {\doibase 10.1088/1361-6455/ac1647} {\bibfield  {journal} {\bibinfo
  {journal} {J. Phys. B: At. Opt. Mol. Phys.}\ }\textbf {\bibinfo {volume}
  {54}},\ \bibinfo {pages} {145303} (\bibinfo {year} {2021})}\BibitemShut
  {NoStop}%
\bibitem [{\citenamefont {Holmstr\"om}(2022)}]{Holmstrom2022}%
  \BibitemOpen
  \bibfield  {author} {\bibinfo {author} {\bibfnamefont {H.}~\bibnamefont
  {Holmstr\"om}},\ }\href {http://lup.lub.lu.se/student-papers/record/9098603}
  {\bibinfo {type} {Bachelor's thesis}},\ \bibinfo  {school} {Lund University}
  (\bibinfo {year} {2022})\BibitemShut {NoStop}%
\bibitem [{\citenamefont {St\"urmer}\ \emph {et~al.}(2022)\citenamefont
  {St\"urmer}, \citenamefont {Tengstrand},\ and\ \citenamefont
  {Reimann}}]{Sturmer2022}%
  \BibitemOpen
  \bibfield  {author} {\bibinfo {author} {\bibfnamefont {P.}~\bibnamefont
  {St\"urmer}}, \bibinfo {author} {\bibfnamefont {M.~N.}\ \bibnamefont
  {Tengstrand}}, \ and\ \bibinfo {author} {\bibfnamefont {S.~M.}\ \bibnamefont
  {Reimann}},\ }\href {\doibase 10.1103/PhysRevResearch.4.043182} {\bibfield
  {journal} {\bibinfo  {journal} {Phys. Rev. Res.}\ }\textbf {\bibinfo {volume}
  {4}},\ \bibinfo {pages} {043182} (\bibinfo {year} {2022})}\BibitemShut
  {NoStop}%
\bibitem [{\citenamefont {Goldstone}(1961)}]{Goldstone1961}%
  \BibitemOpen
  \bibfield  {author} {\bibinfo {author} {\bibfnamefont {J.}~\bibnamefont
  {Goldstone}},\ }\href {\doibase 10.1007/BF02812722} {\bibfield  {journal}
  {\bibinfo  {journal} {Il Nuovo Cimento (1955-1965)}\ }\textbf {\bibinfo
  {volume} {19}},\ \bibinfo {pages} {154} (\bibinfo {year} {1961})}\BibitemShut
  {NoStop}%
\bibitem [{\citenamefont {Brauneis}\ \emph {et~al.}(2022)\citenamefont
  {Brauneis}, \citenamefont {Backert}, \citenamefont {Mistakidis},
  \citenamefont {Lemeshko}, \citenamefont {Hammer},\ and\ \citenamefont
  {Volosniev}}]{Brauneis2022}%
  \BibitemOpen
  \bibfield  {author} {\bibinfo {author} {\bibfnamefont {F.}~\bibnamefont
  {Brauneis}}, \bibinfo {author} {\bibfnamefont {T.~G.}\ \bibnamefont
  {Backert}}, \bibinfo {author} {\bibfnamefont {S.~I.}\ \bibnamefont
  {Mistakidis}}, \bibinfo {author} {\bibfnamefont {M.}~\bibnamefont
  {Lemeshko}}, \bibinfo {author} {\bibfnamefont {H.-W.}\ \bibnamefont
  {Hammer}}, \ and\ \bibinfo {author} {\bibfnamefont {A.~G.}\ \bibnamefont
  {Volosniev}},\ }\href {\doibase 10.1088/1367-2630/ac78d8} {\bibfield
  {journal} {\bibinfo  {journal} {New Journal of Physics}\ }\textbf {\bibinfo
  {volume} {24}},\ \bibinfo {pages} {063036} (\bibinfo {year}
  {2022})}\BibitemShut {NoStop}%
\bibitem [{\citenamefont {Bighin}\ \emph {et~al.}(2022)\citenamefont {Bighin},
  \citenamefont {Burchianti}, \citenamefont {Minardi},\ and\ \citenamefont
  {Macr\`{\i}}}]{Bighin2022}%
  \BibitemOpen
  \bibfield  {author} {\bibinfo {author} {\bibfnamefont {G.}~\bibnamefont
  {Bighin}}, \bibinfo {author} {\bibfnamefont {A.}~\bibnamefont {Burchianti}},
  \bibinfo {author} {\bibfnamefont {F.}~\bibnamefont {Minardi}}, \ and\
  \bibinfo {author} {\bibfnamefont {T.}~\bibnamefont {Macr\`{\i}}},\ }\href
  {\doibase 10.1103/PhysRevA.106.023301} {\bibfield  {journal} {\bibinfo
  {journal} {Phys. Rev. A}\ }\textbf {\bibinfo {volume} {106}},\ \bibinfo
  {pages} {023301} (\bibinfo {year} {2022})}\BibitemShut {NoStop}%
\bibitem [{\citenamefont {Epstein}(1926)}]{Epstein1926}%
  \BibitemOpen
  \bibfield  {author} {\bibinfo {author} {\bibfnamefont {P.~S.}\ \bibnamefont
  {Epstein}},\ }\href {\doibase 10.1103/PhysRev.28.695} {\bibfield  {journal}
  {\bibinfo  {journal} {Phys. Rev.}\ }\textbf {\bibinfo {volume} {28}},\
  \bibinfo {pages} {695} (\bibinfo {year} {1926})}\BibitemShut {NoStop}%
\bibitem [{\citenamefont {Nesbet}(1955)}]{Nesbet1955}%
  \BibitemOpen
  \bibfield  {author} {\bibinfo {author} {\bibfnamefont {R.}~\bibnamefont
  {Nesbet}},\ }\href {\doibase 10.1098/rspa.1955.0134} {\bibfield  {journal}
  {\bibinfo  {journal} {Proc. Royal Soc.  London A:
  Math. Phys. Sci.}\ }\textbf {\bibinfo {volume} {230}},\
  \bibinfo {pages} {312} (\bibinfo {year} {1955})}\BibitemShut {NoStop}%
\bibitem [{\citenamefont {Tubman}\ \emph {et~al.}(2020)\citenamefont {Tubman},
  \citenamefont {Freeman}, \citenamefont {Levine}, \citenamefont {Hait},
  \citenamefont {Head-Gordon},\ and\ \citenamefont {Whaley}}]{Tubman2020}%
  \BibitemOpen
  \bibfield  {author} {\bibinfo {author} {\bibfnamefont {N.~M.}\ \bibnamefont
  {Tubman}}, \bibinfo {author} {\bibfnamefont {C.~D.}\ \bibnamefont {Freeman}},
  \bibinfo {author} {\bibfnamefont {D.~S.}\ \bibnamefont {Levine}}, \bibinfo
  {author} {\bibfnamefont {D.}~\bibnamefont {Hait}}, \bibinfo {author}
  {\bibfnamefont {M.}~\bibnamefont {Head-Gordon}}, \ and\ \bibinfo {author}
  {\bibfnamefont {K.~B.}\ \bibnamefont {Whaley}},\ }\href {\doibase
  10.1021/acs.jctc.8b00536} {\bibfield  {journal} {\bibinfo  {journal} {J. Chem. Th. Comp.}\ }\textbf {\bibinfo {volume} {16}},\
  \bibinfo {pages} {2139} (\bibinfo {year} {2020})}\BibitemShut {NoStop}%
\end{thebibliography}

\end{document}